\documentclass[aps,preprint]{revtex4}

\usepackage{graphicx}

\newcommand{\be}{\begin{equation}}
\newcommand{\ee}{\end{equation}}

\begin{document}
\title{{\bf  Spin Polarized Asymmetric Nuclear Matter and  Neutron Star Matter
Within the Lowest Order Constrained Variational Method}}

\author{{\bf G.H. Bordbar$^{a,b}$ \footnote{Corresponding author} \footnote{E-mail :
bordbar@physics.susc.ac.ir}} and {\bf M. Bigdeli$^{a,c}$}}

 \affiliation{
$^a$Department of Physics, Shiraz University,
Shiraz 71454, Iran\footnote{Permanent address},\\
$^b$Research Institute for Astronomy and Astrophysics of Maragha,\\
P.O. Box 55134-441, Maragha, Iran\\
and\\
 $^c$Department of Physics, Zanjan University, Zanjan, Iran}

%%%%%%%%%%%%%%%%%%%%%%%%%%%%%%%%%%%%%%%%%%%%%%%%%%%%%%%%%%%%%%%%%%%%%%%%%

%%%%%%%%%%%%%%%%%%%%%%%%%%%%%%%%%%%%%%%%%%%%%%%%%%%%%%%%%%%%%%%%%%%%

\begin{abstract}

In this paper, we calculate properties of the spin polarized
asymmetrical nuclear matter and neutron star matter, using the
lowest order constrained variational (LOCV) method with the
$AV_{18}$, $Reid93$, $UV_{14}$ and $AV_{14}$ potentials. According
to our results, the spontaneous phase transition to a
ferromagnetic state in the asymmetrical nuclear matter as well as
neutron star matter do not occur.
\end{abstract}

\maketitle
%%%%%%%%%%%%%%%%%%%%%%%%%%%%%%%%%%%%%%%%%%%%%%%%%%%%%%%
\newpage
\section{Introduction}
The magnetic properties of polarized neutron matter, nuclear
matter and neutron star matter have crucial role for studying the
possible onset of ferromagnetic transition in the neutron star
core. The most interesting and stimulating mechanisms that have
been suggested for magnetic source of pulsars, is the possible
existence of a phase transition to a ferromagnetic state at
densities corresponding to the theoretically stable neutron stars
and, therefore, of a ferromagnetic core in the liquid interior of
such compact objects. Pulsars are rapidly rotating neutron stars
with strong surface magnetic fields in the range of $10^{12}
-10^{13}$ Gauss \cite{shap,paci,gold}. Such a possibility has been
studied by several authors using different theoretical approaches
[4-29], but the results are still contradictory. In most
calculations, neutron star matter is approximated by pure neutron
matter, as proposed just after the discovery of pulsars. Vidana et
al. \cite{vida} and Vidana and Bombaci \cite{vidab} have
considered properties of spin polarized neutron matter and
polarized asymmetrical nuclear matter using the
Brueckner-Hartree-Fock (BHF) approximation by employing three
realistic nucleon- nucleon interactions, Nijmegen II, Reid93 and
NSC97e respectively. Zuo et al. \cite{zls} have also obtained
properties of spin polarized neutron and symmetric nuclear matter
using same method with $AV_{18}$ potential. The results of those
calculations show no indication of ferromagnetic transition at any
density for neutron and asymmetrical nuclear matter. Fantoni et
al. \cite{fanto} have calculated spin susceptibility of neutron
matter using the Auxiliary Field Diffusion Monte Carlo (AFDMC)
method employing the AU6 + UIX three-body potential, and have
found that the magnetic susceptibility of neutron matter shows a
strong reduction of about a factor 3 with respect to its Fermi gas
value. Baldo et al. \cite{bgls}, Akmal et al. \cite{apr} and
Engvik et al. \cite{ehmmp} have considered  properties of  neutron
matter with
 $AV_{18}$ potential using BHF approximation both for continuous choice
(BHFC) and standard choice (BHFG), variational chain summation
(VCS) method and lowest order Brueckner (LOB) respectively. On the
other hand some calculations, like those of Brownell and Callaway
\cite{brown} and Rice \cite{rice} considered a hard sphere gas
model and showed that neutron matter becomes ferromagnetic at
$k_{F} \approx 2.3 fm^{-1}$. Silverstein \cite{silv} and
{\O}stgaard \cite{ostga} found that the inclusion of long range
attraction significantly increased the ferromagnetic transition
density (e.g., {\O}stgaard predicted the transition to occur at
$k_F \approx 4.1 fm^{-1}$ using a simple central potential with
hard core only for the singlet spin states). Clark \cite{clark}
and Pearson and Saunier \cite{pear} calculated the magnetic
susceptibility for low densities $(k_{F} \leq 2 fm^{-1})$ using
more realistic interactions. Pandharipande et al. \cite{pandh},
using the Reid soft-core potential, performed a variational
calculation arriving to the conclusion that such a transition was
not to be expected for $k_{F} \leq5 fm^{-1}$. Early calculations
of the magnetic susceptibility within the Brueckner theory were
performed by B\"{a}ckmann and K\"{a}llman \cite{backm} employing
the Reid soft-core potential, and results from a correlated basis
function calculation were obtained by Jackson et al. \cite{jack}
with the Reid v6 interaction. A different point of view was
followed by Vidaurre et al. \cite{vida}, who employed
neutron-neutron effective interactions of the Skyrme type, finding
the ferromagnetic transition at $k_{F} \approx 1.73 - 1.97
fm^{-1}$. Marcos et al. \cite{marcos} have also studied the spin
stability of dense neutron matter within the relativistic
Dirac-Hartree-Fock approximation with an effective nucleon-meson
Lagrangian, predicting the ferromagnetic transition at several
times nuclear matter saturation density. This transition could
have important consequences for the evolution of a protoneutron
star, in particular for the spin correlations in the medium which
do strongly affect the neutrino cross section and the neutrino
mean free path inside the star \cite{navarro}.

In our pervious works, the properties of unpolarized asymmetrical
nuclear matter have been considered by us  using the lowest order
constrained variational (LOCV) method \cite{bordb} with the
$AV_{18}$ potential \cite{wiring}. We have found that the energy
per particle of nuclear matter alter monotonically and linearly by
quadratic asymmetrical parameter.

Recently, we have computed the properties of polarized neutron
matter\cite{bordbig} and polarized symmetrical nuclear
matter\cite{bordbig2} such as total energy, magnetic
susceptibility, pressure, etc using the microscopic calculations
employing LOCV method with the $AV_{18}$ potential. We have also
concluded that the spontaneous phase transition to a ferromagnetic
state in the neutron and asymmetrical nuclear matter does not
occur. In this work, we intend to calculate the properties of spin
polarized asymmetrical nuclear matter and neutron star matter
using the LOCV method employing the $AV_{18}$ \cite{wiring},
$Reid93$ \cite{R93}, $UV_{14}$ \cite{UV14} and $AV_{14}$
\cite{AV14} potentials.

%-----------------------------------------------------------
\section{LOCV Method}

The LOCV method which was developed several years ago is a useful
tool for the determination of the properties of neutron, nuclear
and asymmetric nuclear matter at zero and finite temperature. The
LOCV method is a fully self-consistent formalism and it does not
bring any free parameters into calculation. It employs a
normalization constraint to keep the higher order term as small as
possible. The functional minimization procedure represents an
enormous computational simplification over unconstrained methods
that attempt to go beyond lowest order \cite{bordb,borda,owen}.

In this method we consider a trial many-body wave function of the
form
\begin{eqnarray}
     \psi=F\phi,
 \end{eqnarray}
where $\phi$ is the uncorrelated ground state wave function
(simply the Slater determinant of plane waves) of $A$ independent
nucleon and $F=F(1\cdots A)$ is an appropriate A-body correlation
operator which can be replaced by a Jastrow form i.e.,
\begin{eqnarray}
    F=S\prod _{i>j}f(ij),
 \end{eqnarray}
in which S is a symmetrizing operator. We consider a cluster
expansion of the energy functional up to the two-body term,
 \begin{eqnarray}
           E([f])=\frac{1}{A}\frac{\langle\psi|H\psi\rangle}
           {\langle\psi|\psi\rangle}=E _{1}+E _{2}\cdot
 \end{eqnarray}
The one-body term $E _{1}$ is total kinetic energy of the system.
 The two-body energy $E_{2}$ is
\begin{eqnarray}\label{ener222}
    E_{2}&=&\frac{1}{2A}\sum_{ij} \langle ij\left| \nu(12)\right|
    ij-ji\rangle,
 \end{eqnarray}
where \\ $\nu(12)=-\frac{\hbar^{2}}{2m}[f(12),[\nabla
_{12}^{2},f(12)]]+f(12)V(12)f(12)$, $f(12)$ and $V(12)$ are the
two-body correlation and potential. For the two-body correlation
function, $f(12)$, we consider the following form
\cite{borda,bordb}:
\begin{eqnarray}
f(12)&=&\sum^3_{k=1}f^{(k)}(12)O^{(k)}(12),
\end{eqnarray}
where, the operators $O^{(k)}(12)$ are given by
\begin{eqnarray}
O^{(k=1-3)}(12)&=&1,\ (\frac{2}{3}+\frac{1}{6}S_{12}),\
(\frac{1}{3}-\frac{1}{6}S_{12}),
\end{eqnarray}
and $S_{12}$ is the tensor operator.
\begin{eqnarray}
S_{12}&=&3(\bf{\sigma_{1}}.\hat{r})(\bf{\sigma_{2}}.\hat{r})-\bf{\sigma_{1}}.\bf{\sigma_{2}}
\end{eqnarray}
%%%%%%%%%%%%%%%%%%%%%%%%%%%%%%%%%%%%%%%%%%%%%%%%%%%%%%%%%%%%%%%%%%%%%%%%%%%%%%%%%%%%%%

Now, we can minimize the two-body energy Eq.(\ref{ener222}), with
respect to the variations in the function ${f_{\alpha}}^{(i)}$ but
subject to the normalization constraint \cite{bordb},

\begin{eqnarray}
       \frac{1}{A}\sum_{ij}\langle ij\left| h_{S_{z}}^{2}
      -f^{2}(12)\right| ij\rangle _{a}=0,
\end{eqnarray}
where in the case of spin polarized neutron matter the function
$h_{S_{z}}(r)$ is defined as
\begin{eqnarray}
      h_{S_{z}}(r)&=&\left[ 1-\frac{9}{\nu}\left( \frac{J_{J}^{2}
     (k_{F}^{i})}{k_{F}^{i}}\right) ^{2}\right] ^{-1/2};\  S_{z}=\pm1
    \nonumber\\
            &=& 1\ \ \ \ \ \ \ \ \ \ \ \ \ \ \ \ \ \ \ \ \ \ \ \ \ \ \
          \ \ \ \ ;\ S_{z}= 0
 \end{eqnarray}
here $\nu$ is the degeneracy of the system. From the minimization
of the two-body cluster energy, we get a set of coupled and
uncoupled differential equations the same as presented in
 Ref. \cite{bordb}. We can get correlation functions by solving the
differential equations and in turn these functions lead to the
two-body energy.
%%%%%%%%%%%%%%%%%%%%%%%%%%%%%%%%%%%%%%%%%%%%%%%%%%%%%%%%%%%%%%%%%%%%%%%%%%%%%%%%%%

\section{Spin Polarized Asymmetrical Nuclear Matter}

Spin polarized asymmetrical nuclear matter is an infinite system
that is composed of spin up and spin down neutrons with densities
$\rho_{n}^{(1)}$ and $\rho_{n}^{(2)}$ respectively, and spin up
and spin down protons with densities $\rho_{p}^{(1)}$ and
$\rho_{p}^{(2)}$ . The total densities for neutrons$(\rho_n)$,
protons $(\rho_{p})$, and nucleons $(\rho)$ are given by:
\begin{eqnarray}
     \rho_{p}=\rho_{p}^{(1)}+\rho_{p}^{(2)},\ \ \ \
     \rho_{n}=\rho_{n}^{(1)}+\rho_{n}^{(2)},
     \ \ \ \ \rho&=&\rho_{p}+\rho_{n}
 \end{eqnarray}
Labels 1 and 2 are used instead of spin up and spin down nucleons,
respectively. One can use the following parameter to identify a
given spin polarized state of the system,
\begin{eqnarray}
      \delta_{p}=\frac{\rho_{p}^{1}-\rho_{p}^{2}}{\rho} , \ \ \
      \delta_{n}=\frac{\rho_{n}^{1}-\rho_{n}^{2}}{\rho}
 \end{eqnarray}
 $\delta_p$ and $\delta_n$ are proton and neutron spin asymmetry parameters,
 respectively.
Total polarization defined as,
\begin{eqnarray}
       \delta&=&\delta_{n} + \delta_{p}
 \end{eqnarray}
For the unpolarized case, we have $\delta_{n}=\delta_{p}=0$.

Asymmetry parameter describes isospin asymmetry of the system and
is defined as,
\begin{eqnarray}
      \beta=\frac{\rho_{n}-\rho_{p}}{\rho}.
 \end{eqnarray}
Pure neutron matter is totally asymmetrical nuclear matter with
$\beta=1$ and symmetrical nuclear matter has $\beta=0$. In the
case of spin polarized asymmetrical nuclear matter, the energy per
particle given by
\begin{eqnarray}\label{asnm}
      E (\rho,\delta,\beta)&=& E_{1}(\rho,\delta,\beta)+
      E_{2}(\rho,\delta,\beta)\cdot
 \end{eqnarray}
where $E_{1}$, the kinetic energy contribution given by,
\begin{eqnarray}
       E_{1}(\rho,\delta,\beta)&=&\frac{3}{20}\frac{\hbar^{2}k_{F}^{2}}{2m}
       \left\{(1+\beta+2\delta_n)^{\frac{5}{3}}+
     (1+\beta-2\delta_n)^{\frac{5}{3}} \right. \nonumber \\&& \left.\ \ \
     \ \ \ \ \ \  +(1-\beta+2\delta_p)^{\frac{5}{3}}
      +(1-\beta-2\delta_p)^{\frac{5}{3}}\right\},
 \end{eqnarray}
 where
$k_{F}=(3/2\pi^{2}\rho)^{1/3}$ is Fermi momentum of unpolarized
symmetrical nuclear matter.
 In expression (\ref{asnm}), the two body energy can be calculated
by the semi-empirical mass formula \cite{bordb},
\begin{eqnarray}\label{ase2}
       E_{2} (\rho,\delta,\beta)&=&
       \beta^{2}{E_{2}}_{neum}+(1-\beta^{2}){E_{2}}_{snucm},
 \end{eqnarray}
 where the powers higher than quadratic are neglected.
 ${E_{2}}_{neum}$ is polarized pure neutron matter potential energy,  and
${E_{2}}_{snucm}$ is polarized symmetrical nuclear matter
potential energy
 that both of them have been determined using the LOCV
method employing a microscopic point of view
\cite{bordbig,bordbig2}.

The energy per particle of the polarized asymmetrical nuclear
matter versus density at different values of the spin polarization
and isospin asymmetry parameter have been shown in Figs. 1-4 for
the $AV_{18}$, $Reid93$, $UV_{14}$ and $AV_{14}$ potentials,
respectively. As it can be seen from these figures, the energy of
polarized asymmetrical nuclear matter for various values of
isospin asymmetry parameters become more repulsive by increasing
the polarization for all relevant densities. There is no crossing
of the energy curves of different polarizations. Reversibly, by
increasing density, the difference between the energy of nuclear
matter at different polarization becomes more sizable. Indeed, the
ground state of the system is that of unpolarized matter in all
ranges of density and isospin asymmetry considered. This behavior
is common for all potentials used in our calculations.

Another quantities that help us to understand ferromagnetic phase
transition, are magnetic susceptibility and Landau parameter. In
what follows, we want to consider these two quantities. The
magnetic susceptibility, $\chi$, which characterizes the response
of a system to the magnetic field and gives a measure of the
energy required to produce a net spin alignment in the direction
of the magnetic field, is defined by
\begin{eqnarray}\label{susc}
\chi =\left( \frac{\partial M}{\partial H}\right) _{H=0},
\end{eqnarray}
where $M$ is the magnetization of the system per unit volume and
$H$ is the magnetic field. We have calculated the magnetic
susceptibility of the polarized asymmetrical nuclear matter in the
ratio ${{\chi}/{\chi_{F}}}$ form. By using the Eq. (\ref{susc})
and some simplification, the ratio of $\chi$ to the magnetic
susceptibility for a degenerate free Fermi gas $\chi_{F}$ can be
written as
\begin{eqnarray}
   \frac{\chi}{\chi_{F}} =\frac{2}{3}\frac{E_{F}
}{\left( \frac{\partial ^{2} E }{\partial \delta ^{2}}\right)
_{\delta =0}}\ ,
\end{eqnarray}
where $E_{F}={\hbar ^{2}k_{F}^{2}}/{2m}$ is the Fermi energy.
After doing some algebra the ratio ${\chi}/{\chi_{F}}$ takes the
form,
\begin{eqnarray}
   \frac{\chi}{\chi_{F}}=\left[1+\frac{
   2\left(\beta^{2}\left\{2^{\frac{2}{3}}({\chi_{F}}/{\chi})_{neum}-
   ({\chi_{F}}/{\chi})_{snucm}\right\}
   +({\chi_{F}}/{\chi})_{snucm}+\beta^{2}(1-2^{\frac{2}{3}})-1\right)}
   {(1+\beta)^{5/3}+(1-\beta)^{5/3}}\right]^{-1},
\end{eqnarray}
where $({\chi}/{\chi_{F}})_{neum}$ and
$({\chi}/{\chi_{F}})_{snucm}$ is magnetic susceptibility of pure
neutron matter \cite{bordbig}, and asymmetry nuclear matter
\cite{bordbig2}, respectively. In Fig. 5, we have plotted the
ratio ${\chi}/{\chi_{F}}$ as a function of density at several
values of the isospin asymmetry for the $AV_{18}$, $Reid93$,
$UV_{14}$ and $AV_{14}$ potentials.  For all used potentials, we
see that the value of the ratio ${\chi}/{\chi_{F}}$ decreases
monotonically with the density, even at high densities. It shows,
there is no magnetic instability for any values of asymmetry
parameter. If such an instability existed, the value of ratio
${\chi}/{\chi_{F}}$ changes suddenly unlike previous treatment.

As it is known, the Landau parameter, $G_{0}$, describes the spin
density fluctuation in the effective interaction. $G_{0}$ is
simply related to the magnetic susceptibility by the relation
\begin{eqnarray}
   \frac{\chi}{\chi_{F}} =\frac{m^{*}
}{1+G_{0}},
\end{eqnarray}
where $m^{*}$ is the effective mass. A magnetic instability would
require $G_{0}<-1$. Our results for the Landau parameter have been
presented in the Fig. 6 for the $AV_{18}$, $Reid93$, $UV_{14}$ and
$AV_{14}$ potentials. It is seen that the value of $G_{0}$ is
always positive and monotonically increases by increasing the
density. This shows that for all used potentials, the spontaneous
phase transition to a ferromagnetic state in the asymmetrical
nuclear matter does not occur.

The equation of state of polarized asymmetrical nuclear matter,
$P(\rho,\beta,\delta)$, can be obtained using
\begin{eqnarray}
      P(\rho,\beta,\delta)= \rho^{2} \frac{\partial E(\rho,\beta,\delta)}{\partial
      \rho}.
 \end{eqnarray}
In Fig. 7, we have shown the pressure of asymmetrical nuclear
matter as a function of density ($\rho$) at different
polarizations for various choice of asymmetry parameter ($\beta$)
with the $AV_{18}$ potential. This figure shows that the equation
of state becomes stiffer by increasing the polarization for all
isospin asymmetry.

%%%%%%%%%%%%%%%%%%%%%%%%%%%%%%%%%%%%%%%%%%%%%%%%%%%%%%%%%%%%%%%%%%%%%%%%%%%%%%%%%%%%%%%%%%%%%%%%%%%%%%%%%%%%%%%%%%%
%-------------------------------------------------------------------------
\section{Polarized Neutron Star Matter}
Indeed the neutron star matter is charge neutral infinite system
that is mixture of asymmetrical nuclear matter and leptons,
specially electrons and muons. The energy per particle of
polarized neutron star matter, $E_{nsm}$ can be written as,
\begin{eqnarray}
       E_{nsm}&=&E +E_{l}\ .
 \end{eqnarray}
where $E$ is the nucleonic energy contribution which given by Eq.
(\ref{asnm}) and $E_{l}$ is leptonic energy contribution obtained
as follows,
\begin{eqnarray}
       E_{l}&=&\sum_{i=e,\ \mu}\  \sum_{k\leq k^{(F)}}
       [(m_{i}c^{2})^{2}+\hbar^{2}c^{2}k^{2}]^{1/2} \ .
 \end{eqnarray}
After some simplification the above equation leads to
\begin{eqnarray}\label{engl}
      E_{l}=\frac{3}{8}\sum_{i=e,\ \mu}\frac{\rho_i}{\rho}
      \frac{m_{i}c^{2}}{{x_{i}^{(F)}}^{3}}\left[
       x_{i}^{(F)}(2{x_{i}^{(F)}}^{2}+1)\sqrt{{x_{i}^{(F)}}^{2}+1}\
       -\sinh^{-1}x_{i}^{(F)}\right],
 \end{eqnarray}
  where $x_{i}^{(F)}=\hbar k_{i}^{(F)}/m_i c$\ .
The conditions of charge neutrality and beta equilibrium impose
the following constraints on the calculation of energy of neutron
star matter \cite{shap},
\begin{eqnarray}
      \mu_{n}&=&\mu_{p}+\mu_{e}\ \ \ \ \ \ \ \mu_{e}=\mu_{\mu}\ .
 \end{eqnarray}
\begin{eqnarray}
      \rho_{p}&=&\rho_{e}+\rho_{\mu}\ .
 \end{eqnarray}
In Fig. 8, we have shown the energy per particle of the neutron
star matter at various values of spin polarization as a function
of density for the $AV_{18}$, $Reid93$, $UV_{14}$ and $AV_{14}$
potentials. The same as Figs. 1-4, this figure have also shown no
phase transition to a ferromagnetic state.

The treatment of magnetic susceptibility and landau parameter of
neutron star matter versus density for different polarization,
have been shown in Figs. 9 and 10, respectively. This treatments
are the same as the treatments of the polarized neutron matter,
symmetrical and asymmetrical nuclear matter, and it does not show
any magnetic instability for the neutron star matter. This
behavior has been observed for all potentials used in our
calculations.

For the $AV_{18}$ potential, the pressure of neutron star matter
have been presented as a function of density $\rho$ at different
polarizations in Fig. 11. We see that the equation of state of
neutron star matter becomes stiffer by increasing the
polarization.

%----------------------------------------------------------------------------------
\section{Summary and Conclusions}
The purpose of this paper was to calculate the properties of the
spin polarized asymmetrical nuclear matter and neutron star matter
employing the lowest order constrained variational technique with
the $AV_{18}$, $Reid93$, $UV_{14}$ and $AV_{14}$ nucleon-nucleon
potentials. After introducing the LOCV method briefly, we
calculated the energy per particle of the polarized asymmetrical
nuclear matter and showed that the force becomes more repulsive by
increasing the polarization for all relevant densities. No
crossing of the energy curves was observed and the ground state of
the system was found to be that of the unpolarized matter in all
ranges of densities and isospin asymmetry. The magnetic
susceptibility was computed and shown to remain almost constant by
increasing the asymmetry parameter. For the polarized neutron star
matter, we showed that there is no magnetic instability and the
equation of state becomes stiffer by increasing the polarization.

%%%%%%%%%%%%%%%%%%%%%%%%%%%%%%%%%%%%%%%%%%%%%%%%%%%%%%%%%%%%%%%%%
\acknowledgements{ This work has been supported financially by
Research Institute for Astronomy and Astrophysics of Maragha. One
os us (G.H. Bordbar) wish to thanks Shiraz University Research
Council.}

%%%%%%%%%%%%%%%%%%%%%%%%%%%%%%%%%%%%%%%%%%%%%%%%%%%%%%%%%%%%%%

%%
%%%%%%%%%%%%%%%%%%%%%%%%%%%%%%%%%%%%%%%%%%%%%%%%%%%%%%%%%%%%%%%%%%
%%%%%%%%%%%%%%%%%%%%%%%%%%%%%%%%%%%%%%%%%%%%%%%%%%%%%%%%%%%%%%%%%%%

%%%%%%%%%%%%%%%%%%%%%%%%%%%%%%%%%%%%%%%%%%%%%%%%%%%%%%%%%%%%%%%%%%%%%%%%%%%%%%%%%%%%%%%%%%%%%%%%%%%%%%%
\newpage
%%%%%%%%%%%%%%%%%%%%%%%%%%%%%%%%%%%%%%%%%%%%%%%%%%%%%%%%%%%%%%%%%%%%%%%%%%%%%%%

%%%%%%%%%%%%%%%%%%%%%%%%%%%%%%%%%%%%%%%%%%%%%%%%%%%%%%%%%%%%%%%%%

\begin{figure}
\includegraphics[height=2.5in]{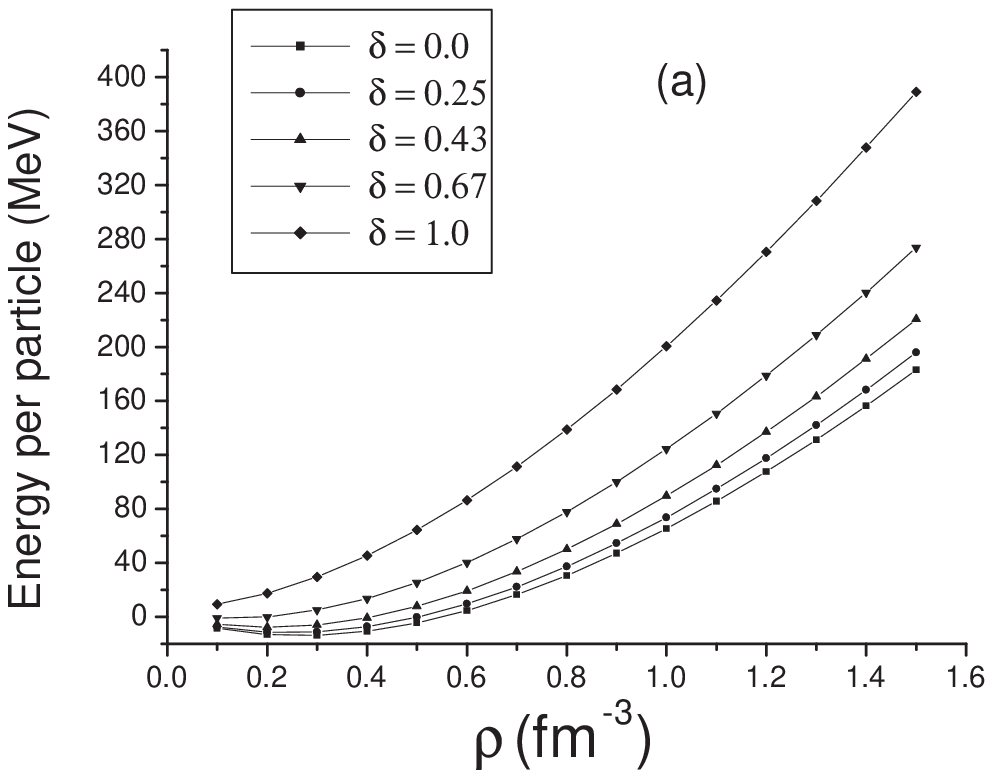}
\includegraphics[height=2.5in]{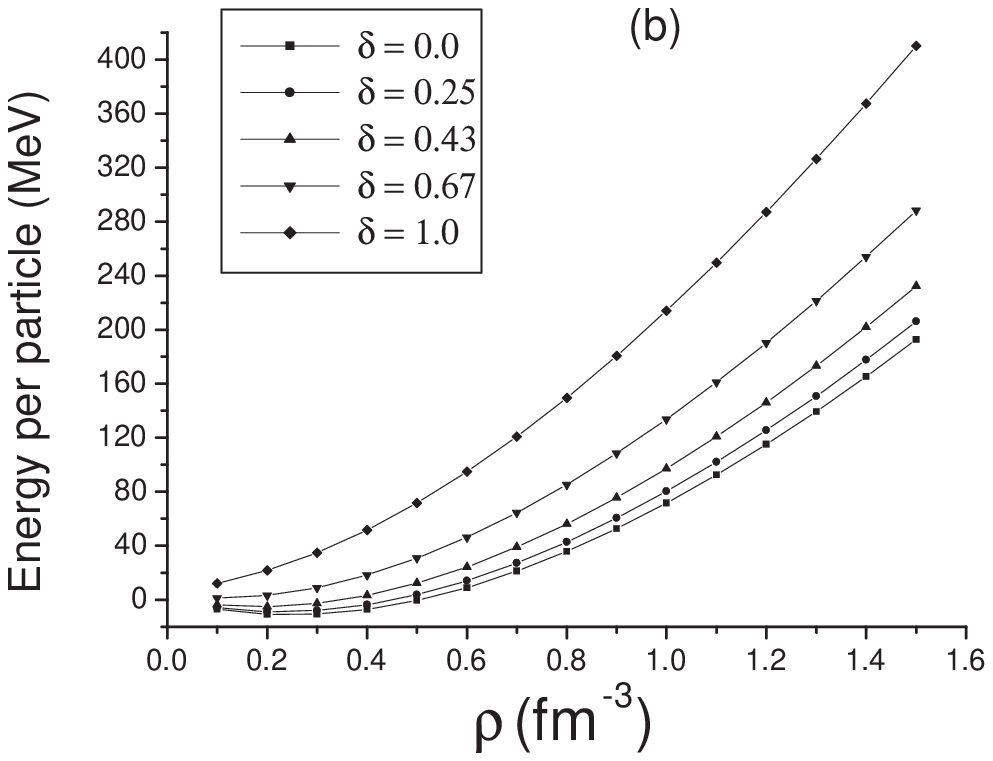}
\includegraphics[height=2.5in]{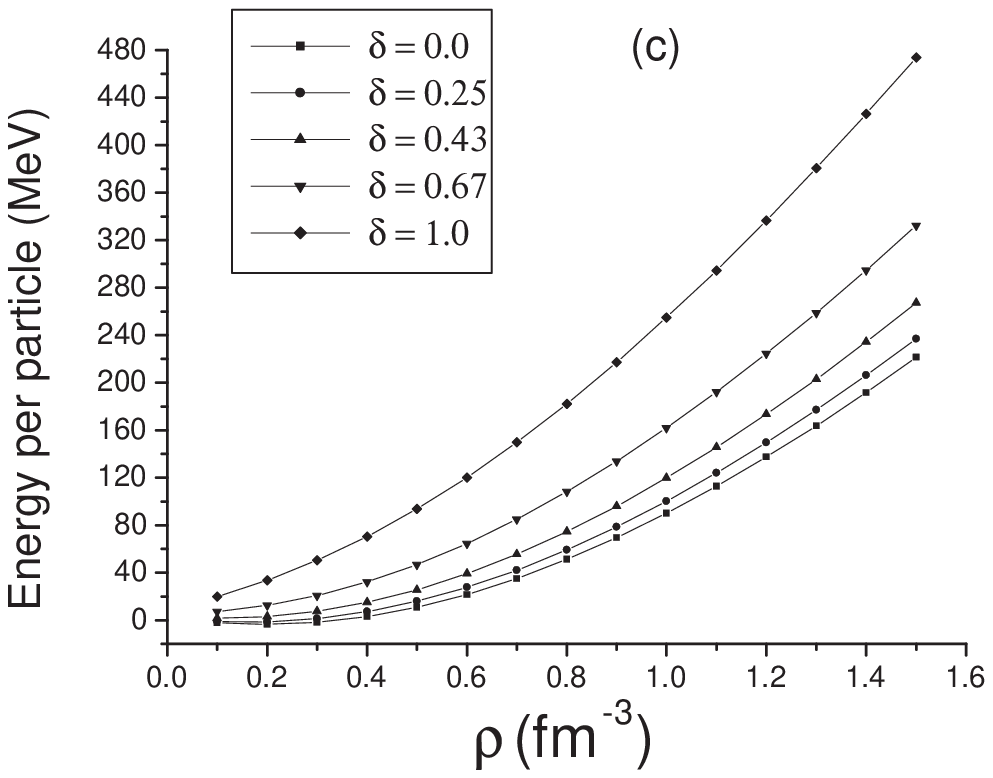}
\includegraphics[height=2.5in]{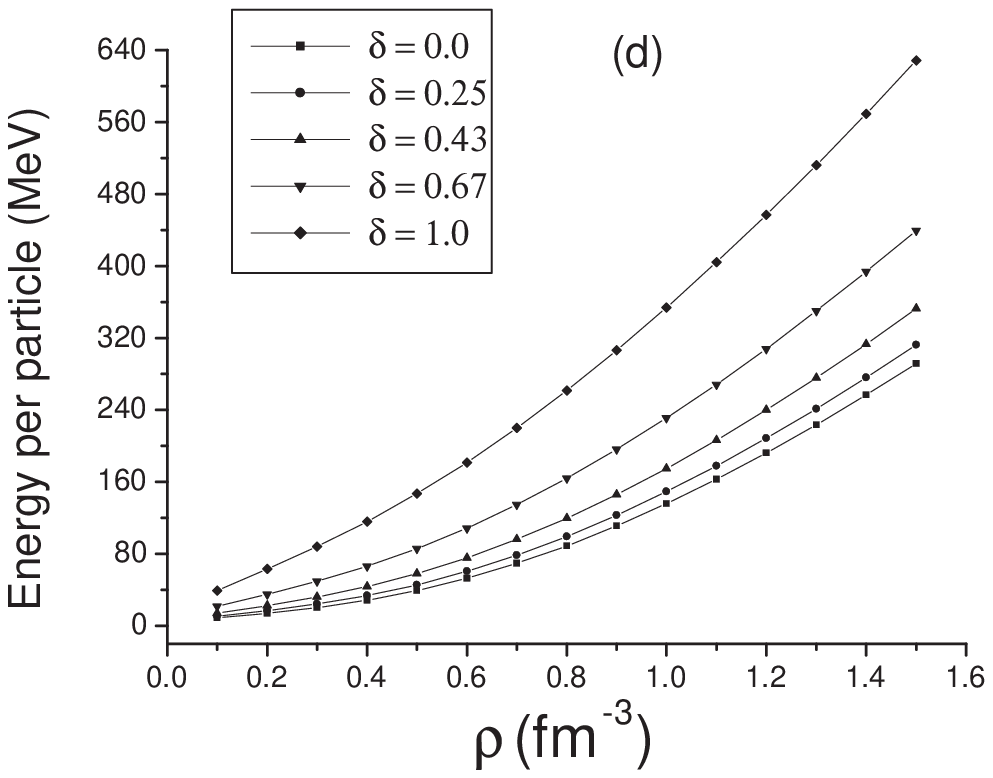}
 \caption{The energy
per particle of the polarized asymmetrical nuclear matter versus
density($\rho$) for different values of the spin polarization
($\delta$) with the $AV_{18}$ potential  at $\beta=0.0$ (a),
$\beta=0.3$ (b), $\beta=0.6$ (c), $\beta=1.0$ (d).}
\end{figure}
%%%%%%%%%%%%%%%%%%%%%%%%%%%%%%%%%%%%%%%%%%%%%%%%%%%%%%%%%%%%%%%%%%%%%%%%%%%%%
\newpage
%%%%%%%%%%%%%%%%%%%%%%%%%%%%%%%%%%%%%%%%%%%%%%%%%%%%%%%%%%%%%%%%%

\begin{figure}
\includegraphics[height=2.5in]{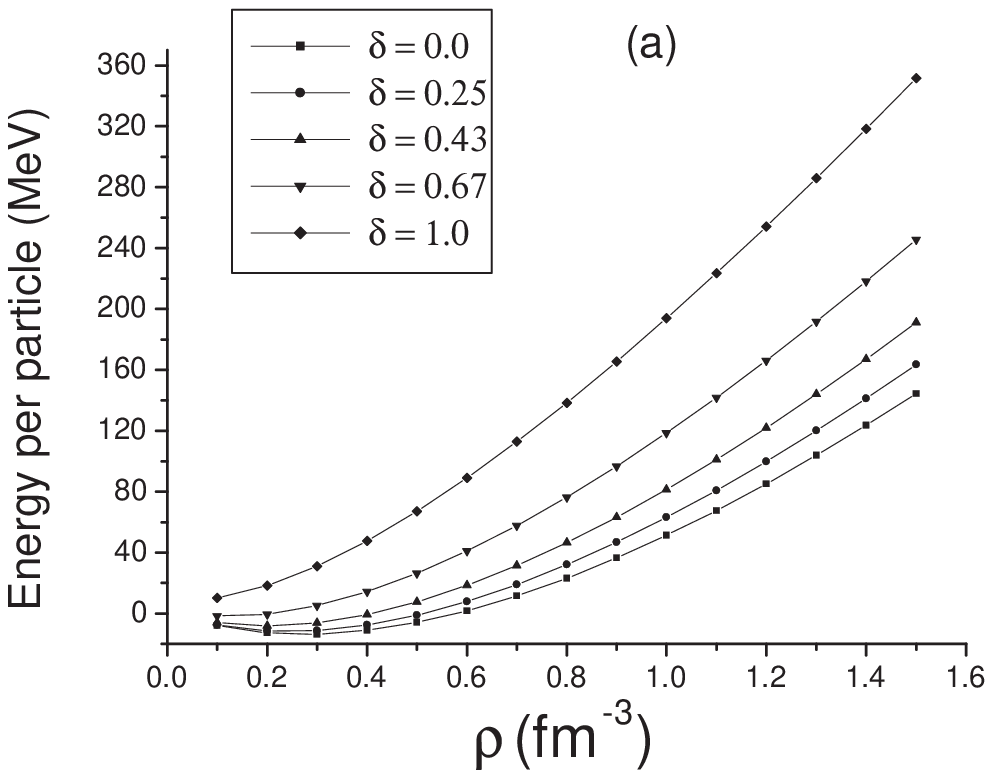}
\includegraphics[height=2.5in]{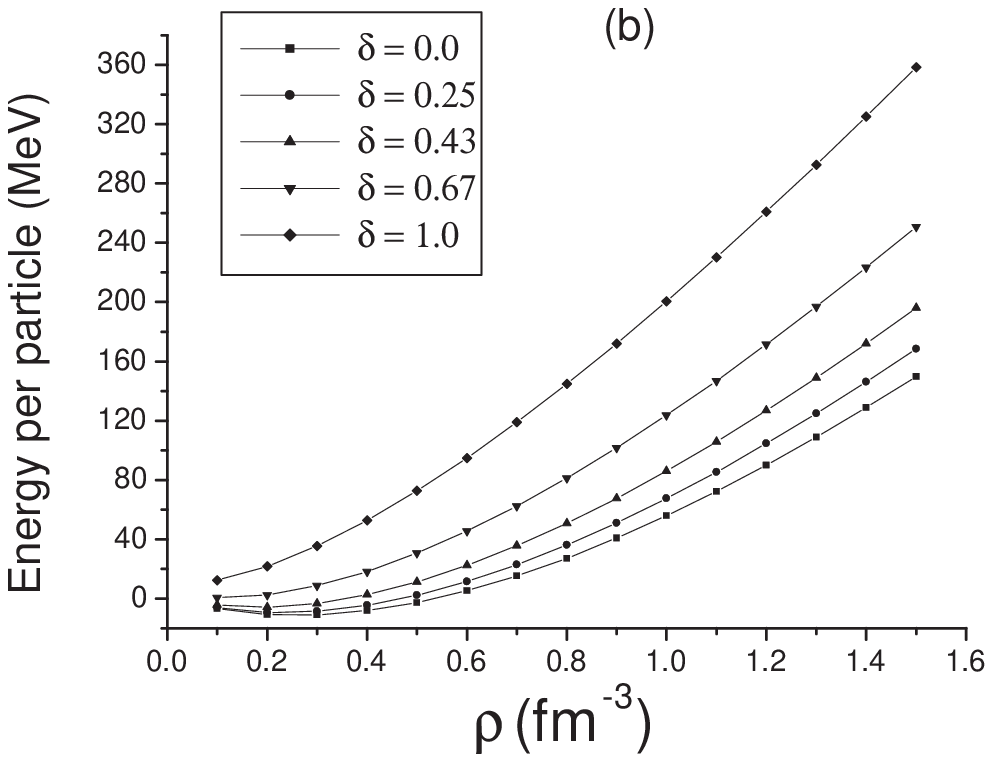}
\includegraphics[height=2.5in]{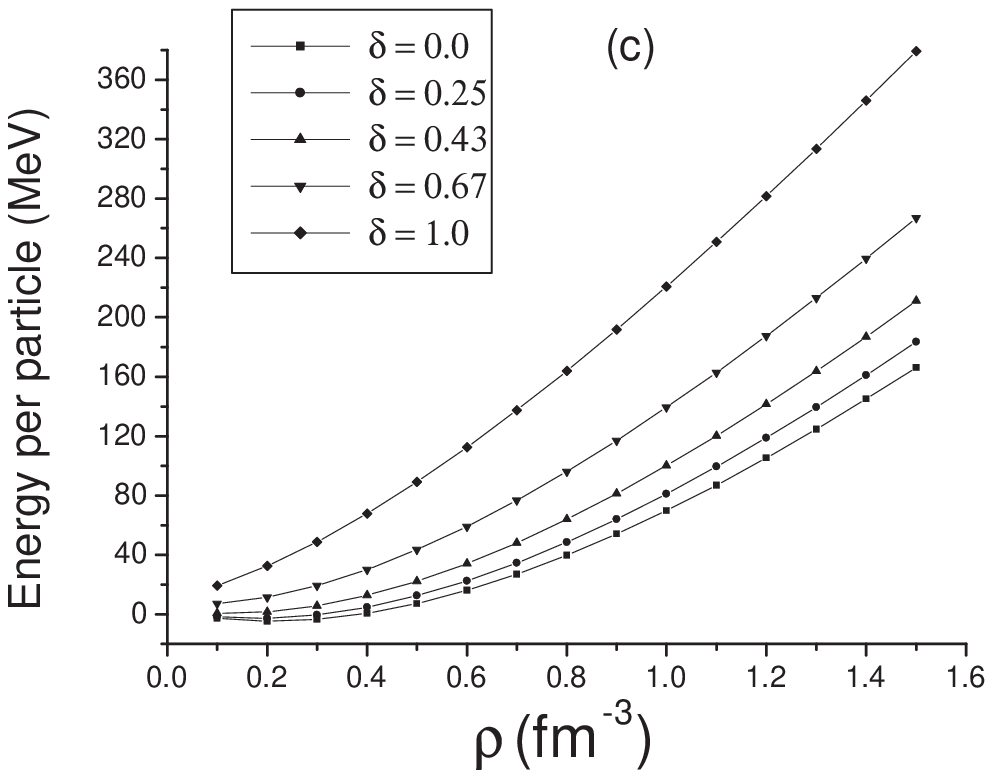}
\includegraphics[height=2.5in]{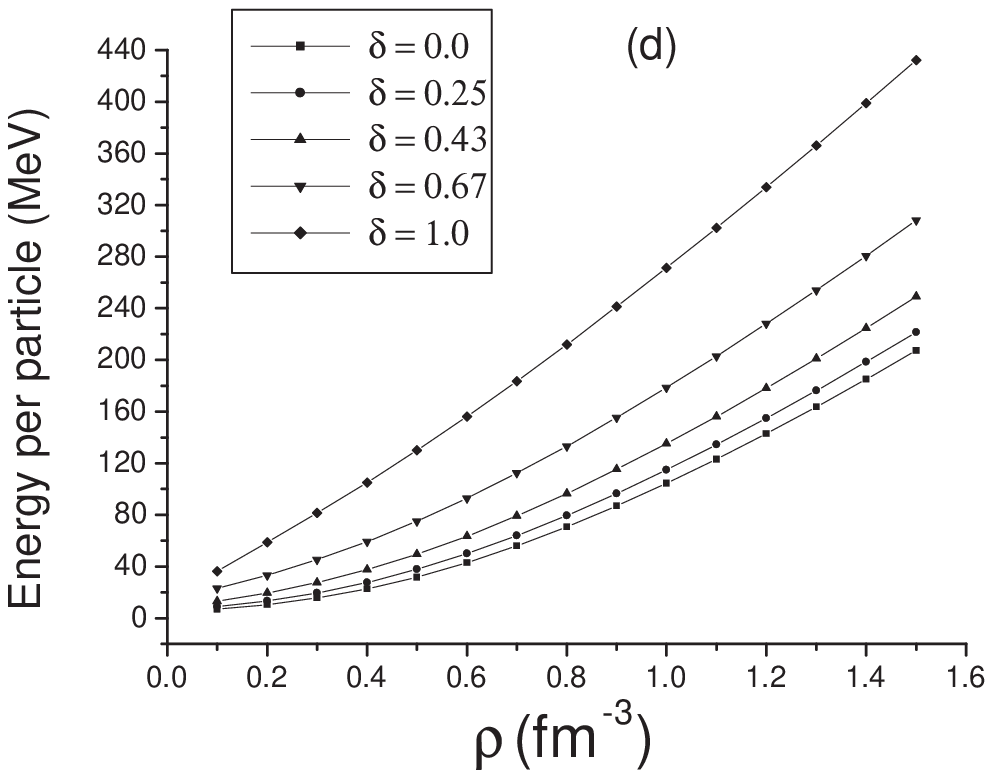}
\caption{The energy per particle of the polarized asymmetrical
nuclear matter versus density($\rho$) for different values of the
spin polarization ($\delta$) with the Reid93 potential at
$\beta=0.0$ (a), $\beta=0.3$ (b), $\beta=0.6$ (c), $\beta=1.0$
(d).}
\end{figure}
%%%%%%%%%%%%%%%%%%%%%%%%%%%%%%%%%%%%%%%%%%%%%%%%%%%%%%%%%%%%%%%%%%%%%%%%%%%%%
%%%%%%%%%%%%%%%%%%%%%%%%%%%%%%%%%%%%%%%%%%%%%%%%%%%%%%%%%%%%%%%%%

\begin{figure}
\includegraphics[height=2.6in]{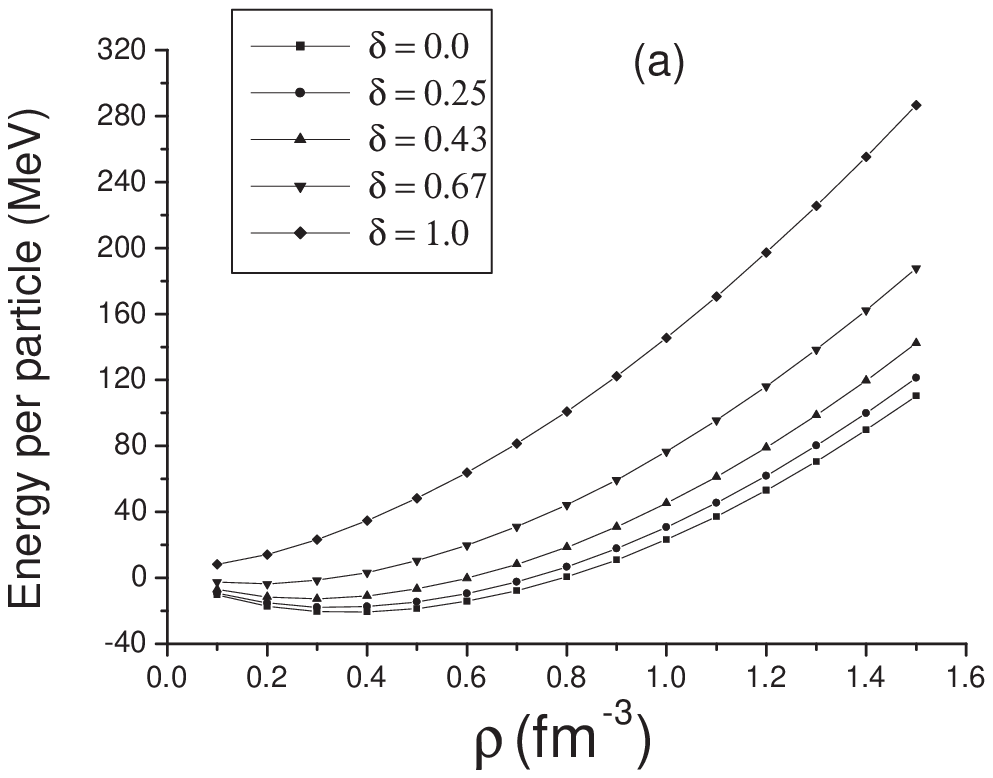}
\includegraphics[height=2.6in]{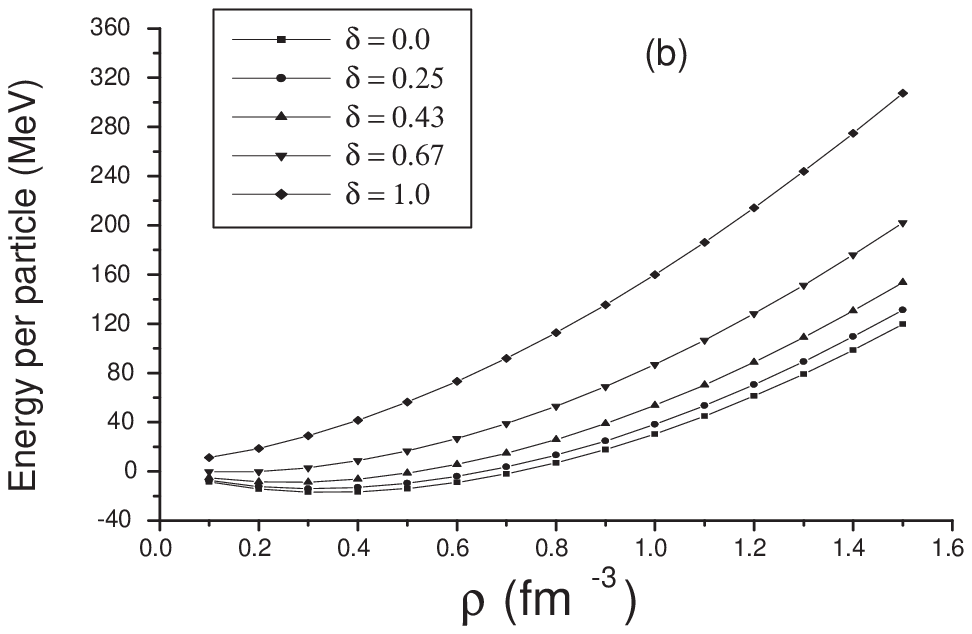}
\includegraphics[height=2.6in]{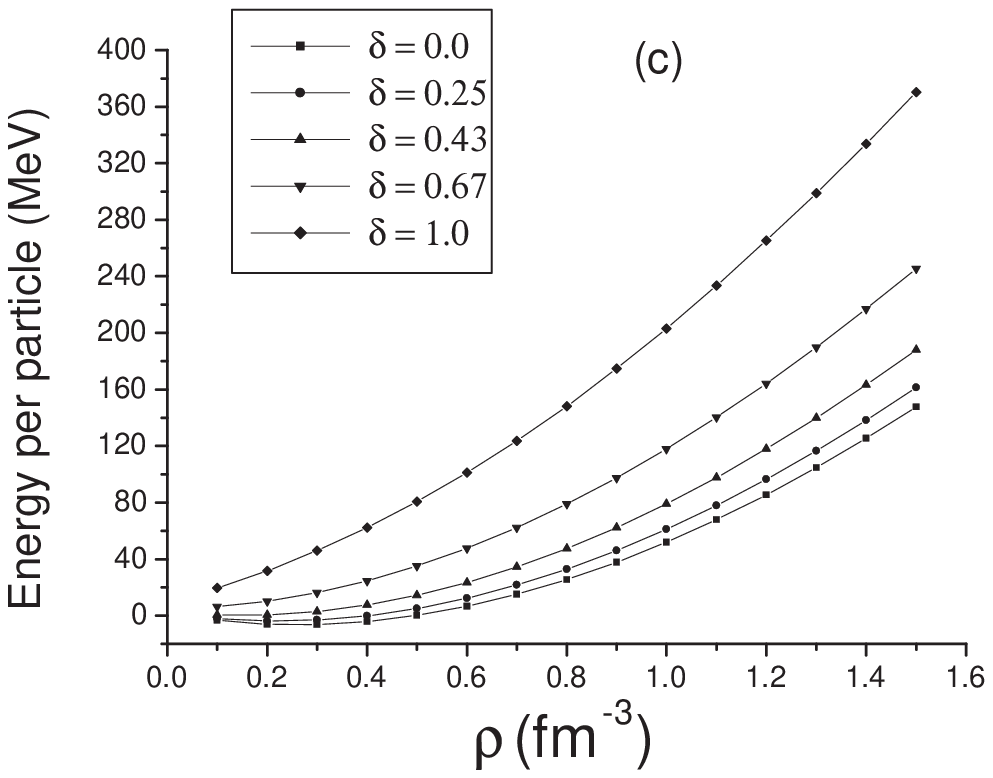}
\includegraphics[height=2.6in]{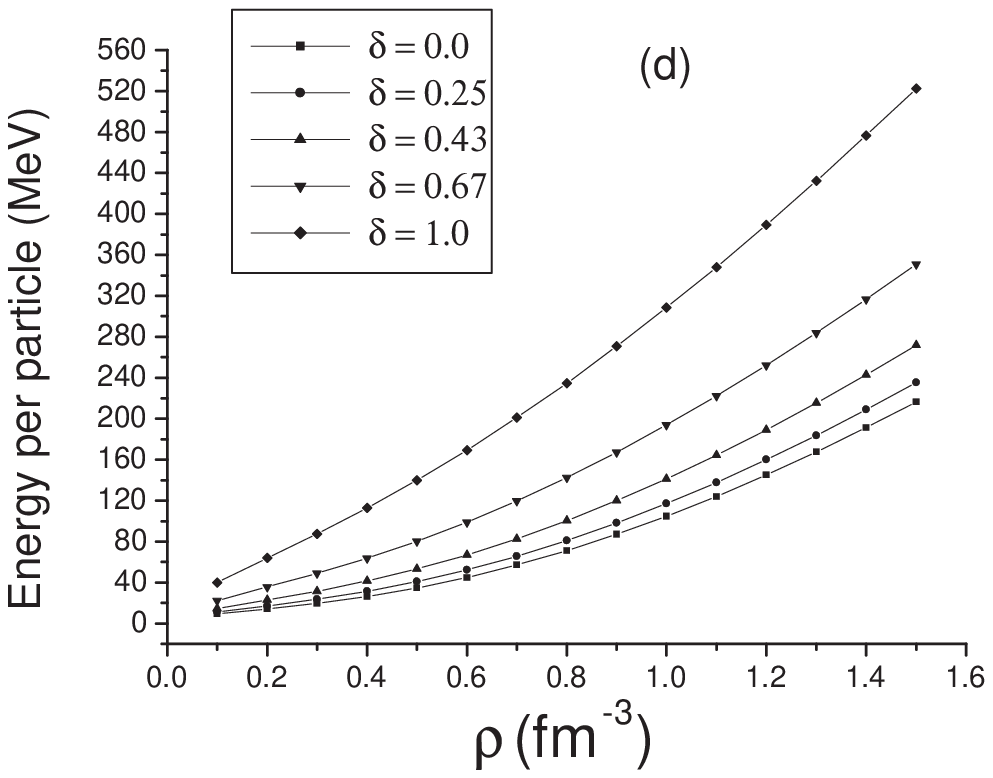}
\caption{The energy per particle of the polarized asymmetrical
nuclear matter versus density($\rho$) for different values of the
spin polarization ($\delta$) with the $UV_{14}$ potential at
$\beta=0.0$ (a), $\beta=0.3$ (b), $\beta=0.6$ (c), $\beta=1.0$
(d).}
\end{figure}
%%%%%%%%%%%%%%%%%%%%%%%%%%%%%%%%%%%%%%%%%%%%%%%%%%%%%%%%%%%%%%%%%%%%%%%%%%%%%
%%%%%%%%%%%%%%%%%%%%%%%%%%%%%%%%%%%%%%%%%%%%%%%%%%%%%%%%%%%%%%%%%

\begin{figure}
\includegraphics[height=2.5in]{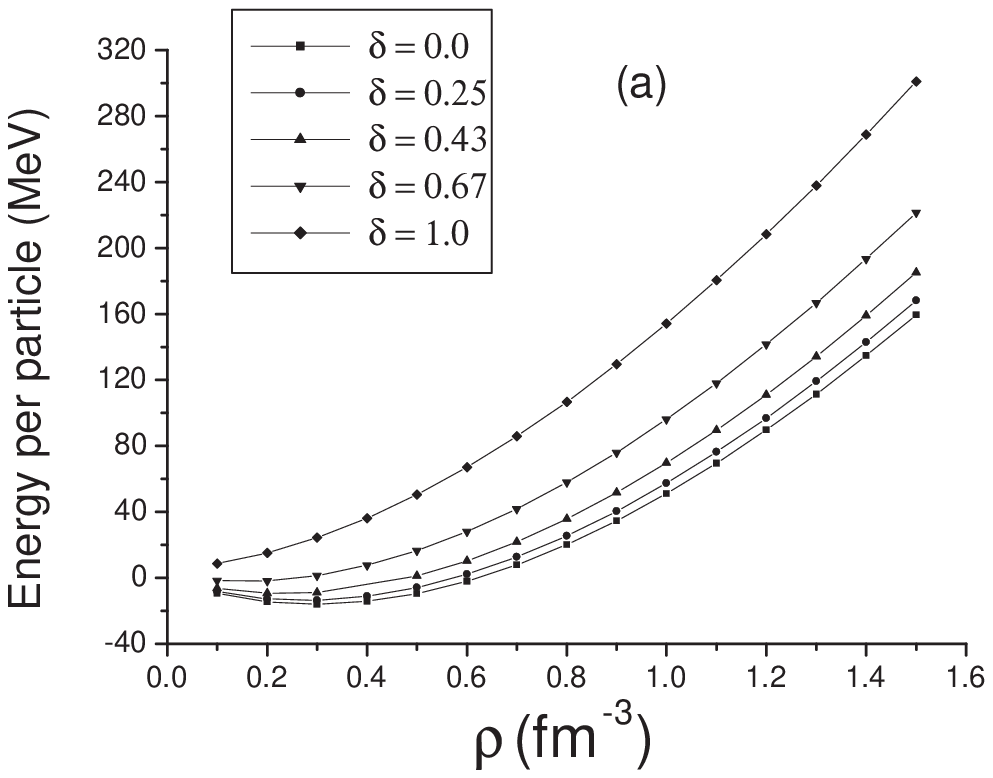}
\includegraphics[height=2.5in]{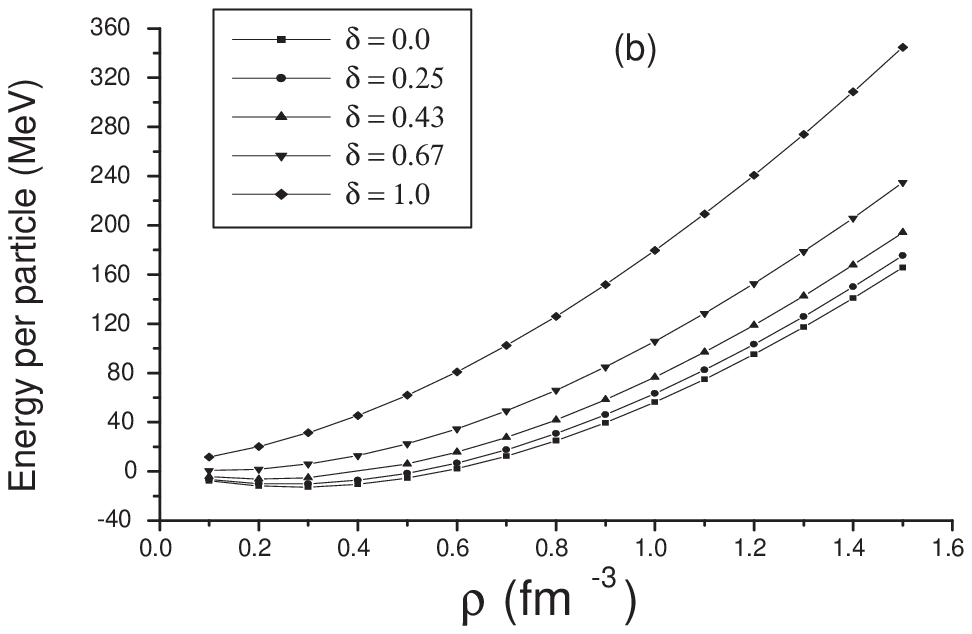}
\includegraphics[height=2.5in]{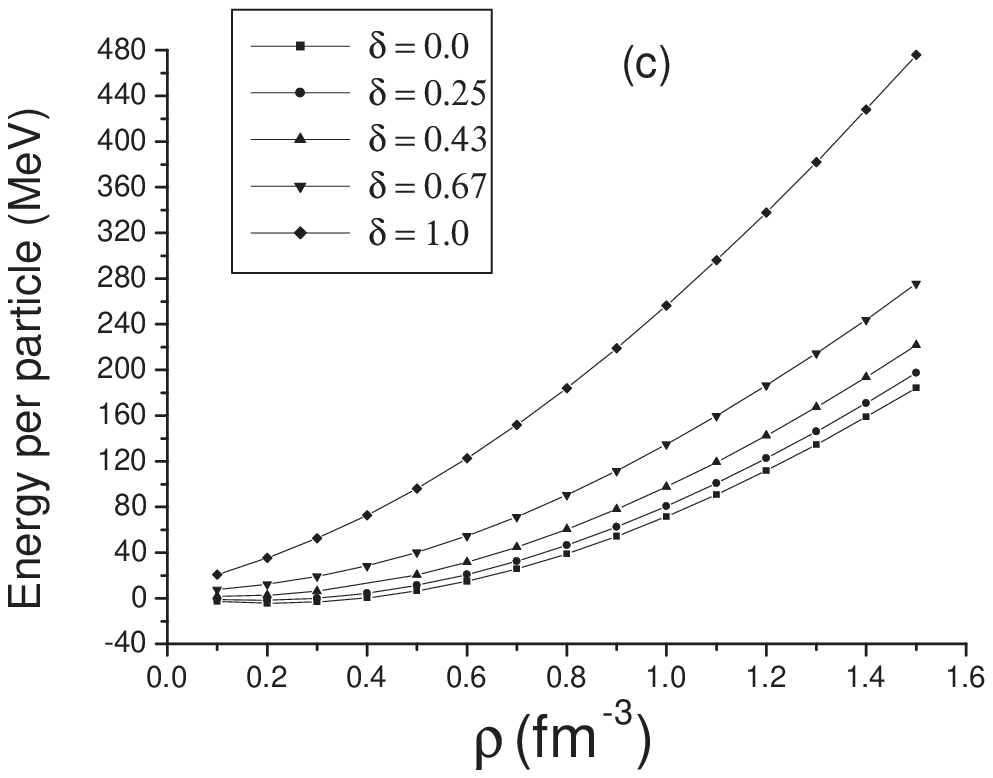}
\includegraphics[height=2.5in]{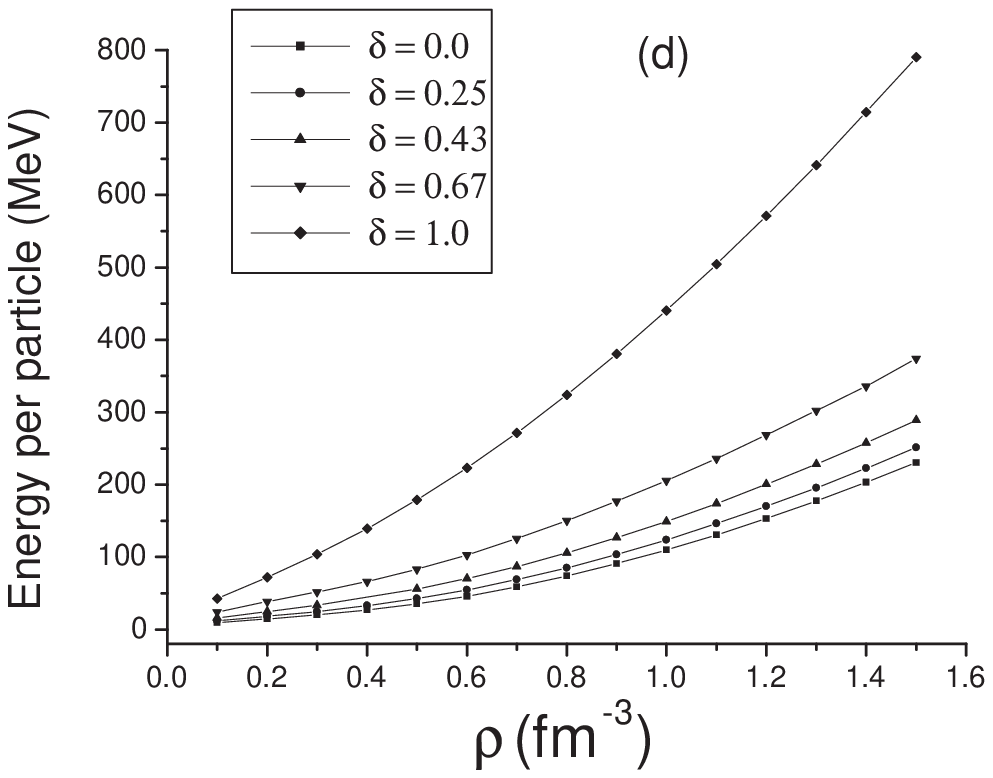}
\caption{The energy per particle of the polarized asymmetrical
nuclear matter versus density($\rho$) for different values of the
spin polarization ($\delta$) with the $AV_{14}$ potential at
$\beta=0.0$ (a), $\beta=0.3$ (b), $\beta=0.6$ (c), $\beta=1.0$
(d).}
\end{figure}
%%%%%%%%%%%%%%%%%%%%%%%%%%%%%%%%%%%%%%%%%%%%%%%%%%%%%%%%%%%%%%%%%%%%%%%%%%%%%
\newpage

\begin{figure}
\includegraphics[height=2.5in]{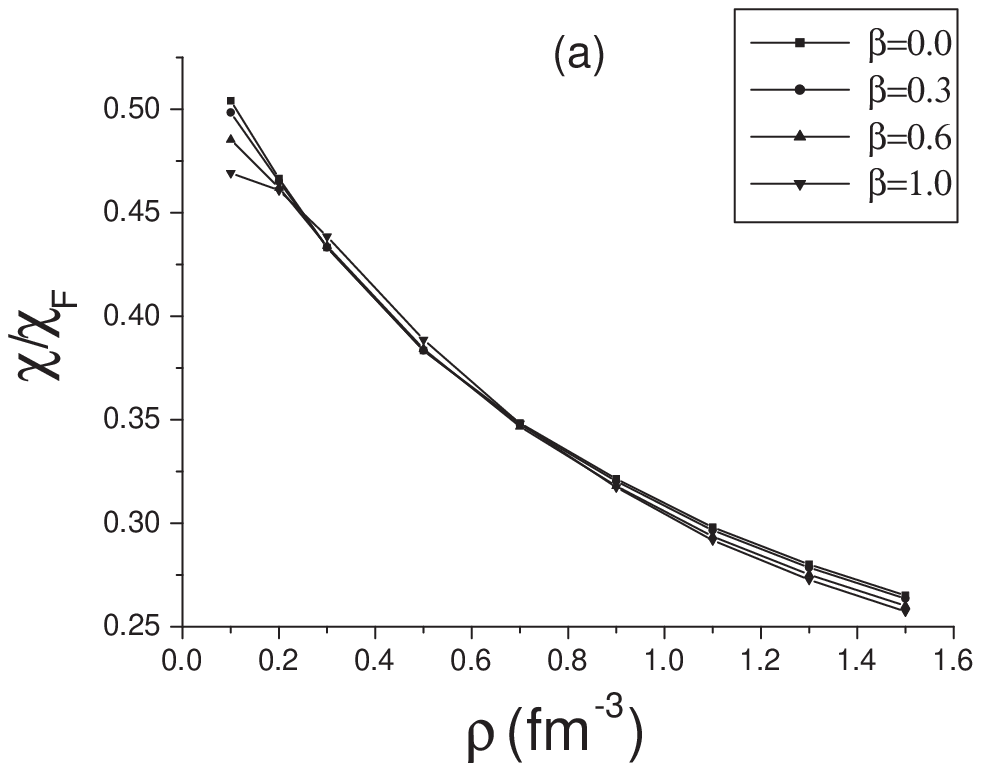}
\includegraphics[height=2.5in]{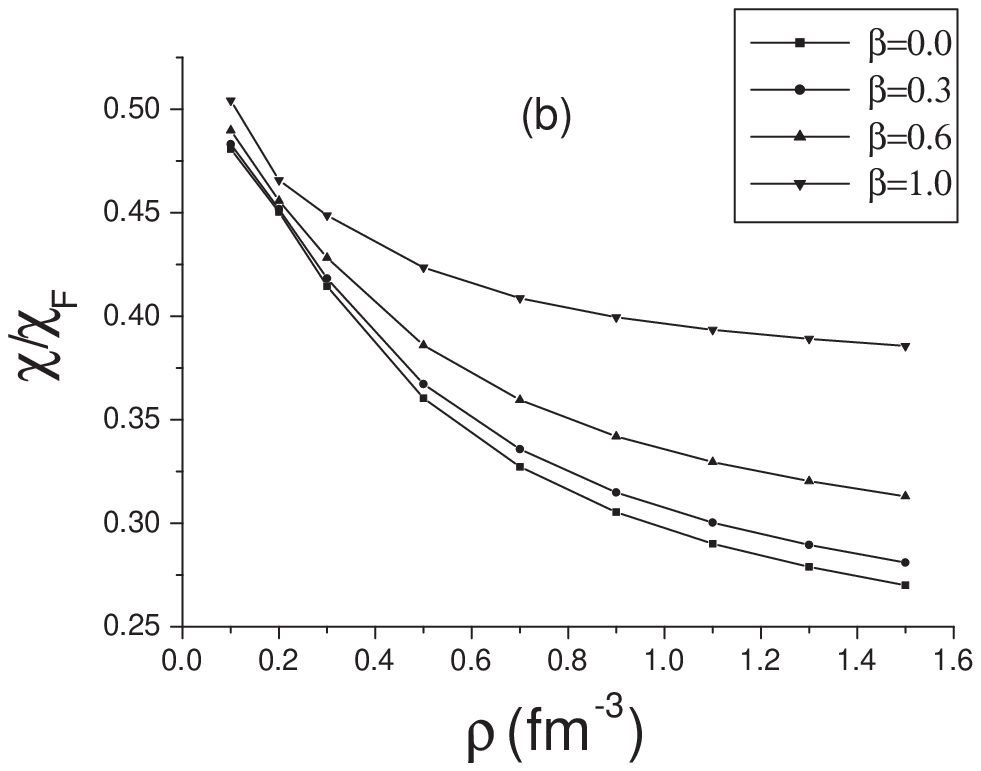}
\includegraphics[height=2.5in]{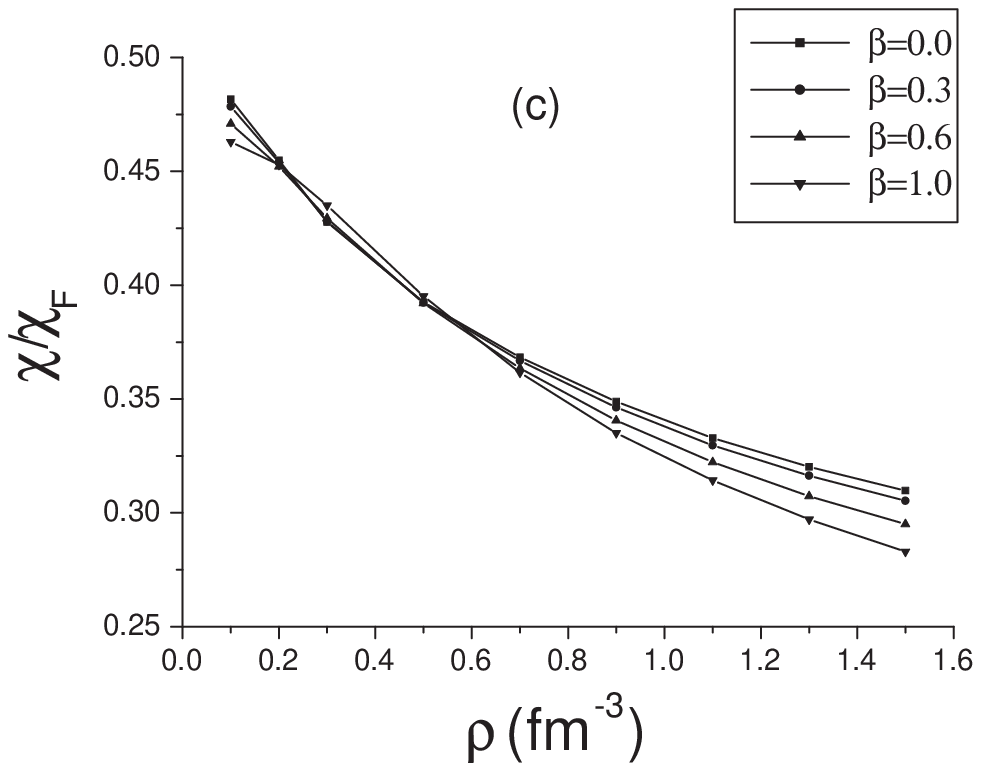}
\includegraphics[height=2.5in]{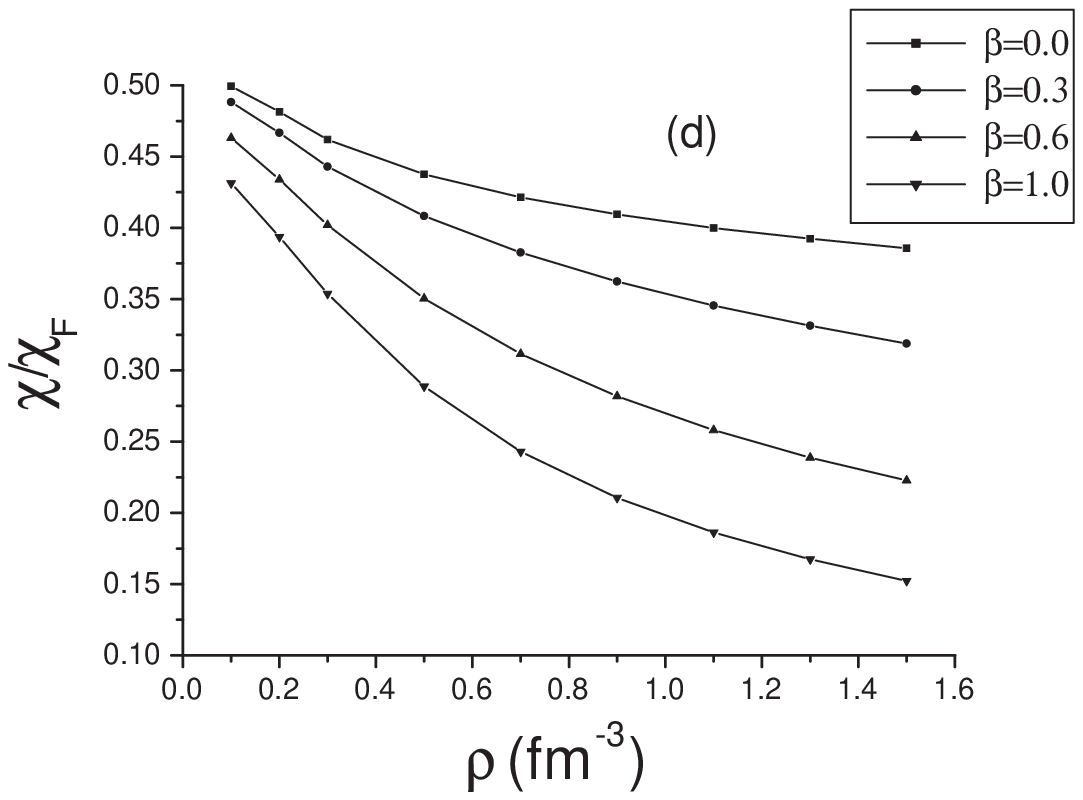}
\caption{The magnetic susceptibility of the polarized asymmetrical
nuclear matter as the function of density ($\rho$) for different
values of asymmetry $(\beta)$ with the $AV_{18}$ (a), Reid93 (b),
$UV_{14}$ (c) and $AV_{14}$ (d) potentials. } \label{sus(den)}
\end{figure}

%%%%%%%%%%%%%%%%%%%%%%%%%%%%%%%%%%%%%%%%%%%%%%%%%%%%%%%%%%%%%%%%%%%%%%%%%%%%%%%
%%%%%%%%%%%%%%%%%%%%%%%%%%%%%%%%%%%%%%%%%%%%%%%%%%%%%%%%%%%%%%%%%%%%%%%%%%%%%
\newpage

\begin{figure}
\includegraphics[height=2.5in]{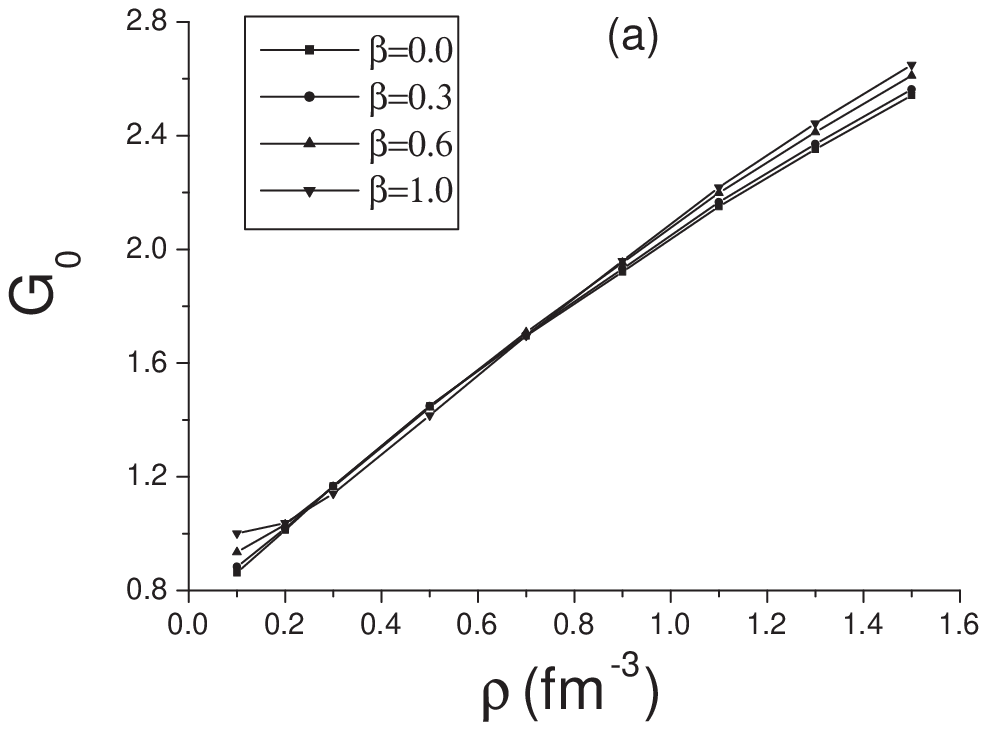}
\includegraphics[height=2.5in]{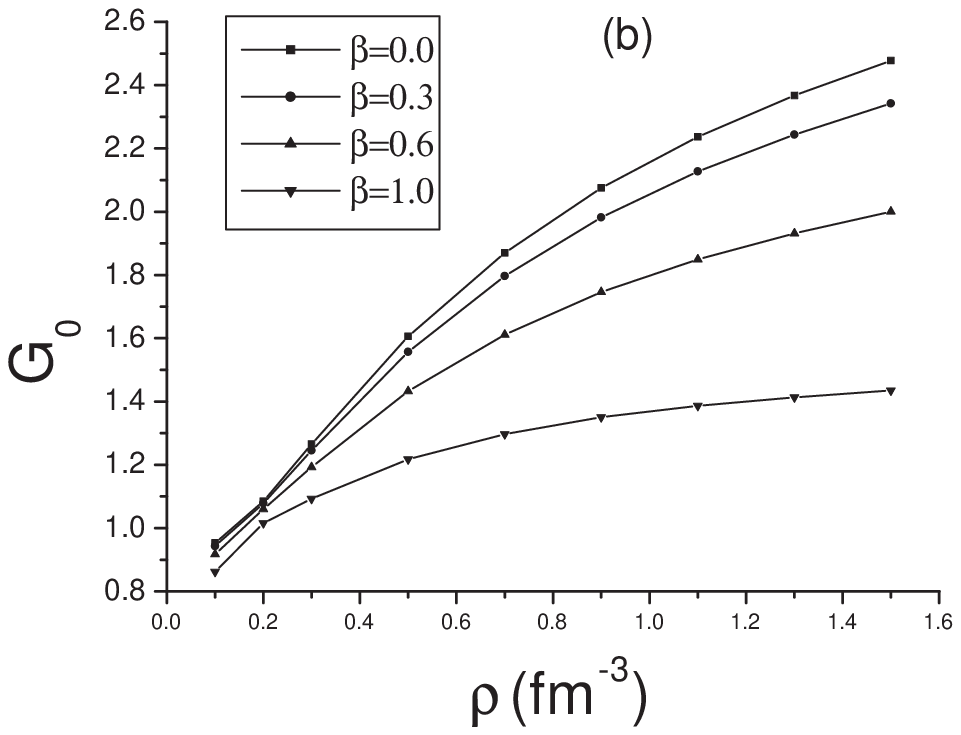}
\includegraphics[height=2.5in]{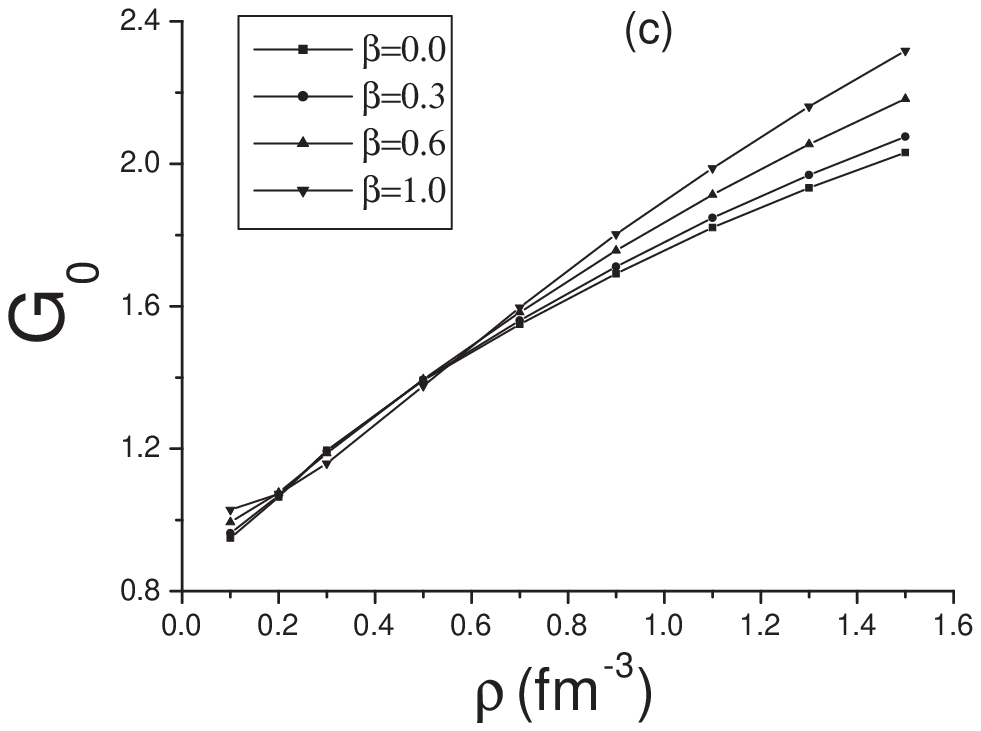}
\includegraphics[height=2.5in]{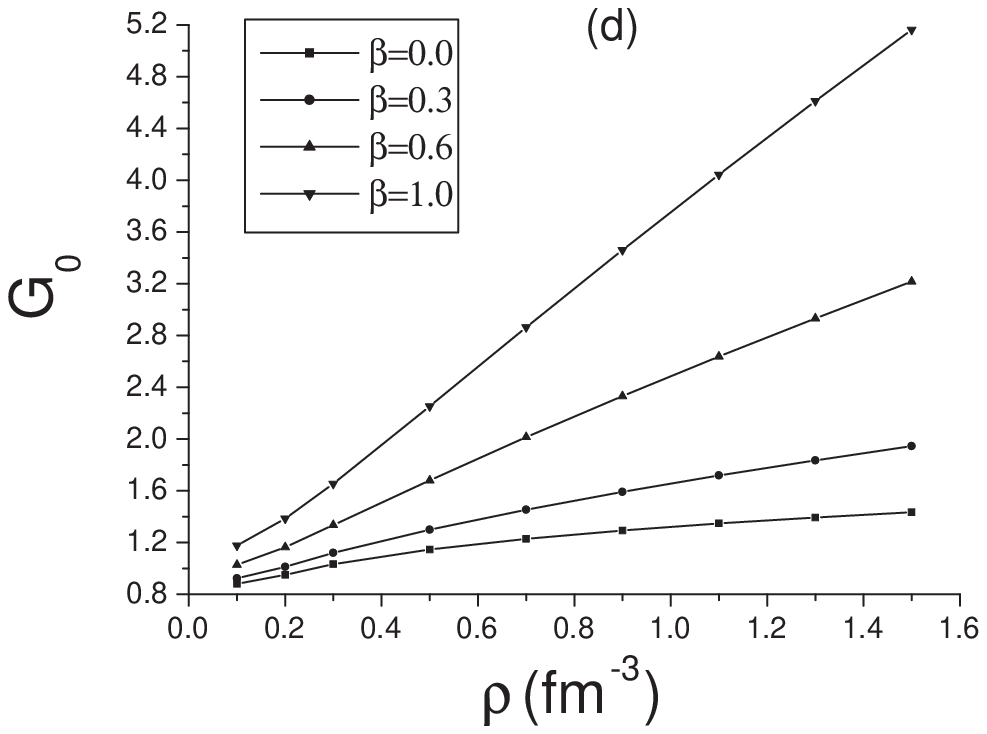}
\caption{The Landau parameter of the polarized asymmetrical
nuclear matter as the function of density ($\rho$) for different
values of asymmetry $(\beta)$ with the $AV_{18}$ (a), Reid93 (b),
$UV_{14}$ (c) and $AV_{14}$ (d) potentials.} \label{G(den)}
\end{figure}
\newpage

%%%%%%%%%%%%%%%%%%%%%%%%%%%%%%%%%%%%%%%%%%%%%%%%%%%%%%%%%%%%%%%%%%%%%%%%%%%%%%%%
\newpage

\begin{figure}
\includegraphics[height=2.5in]{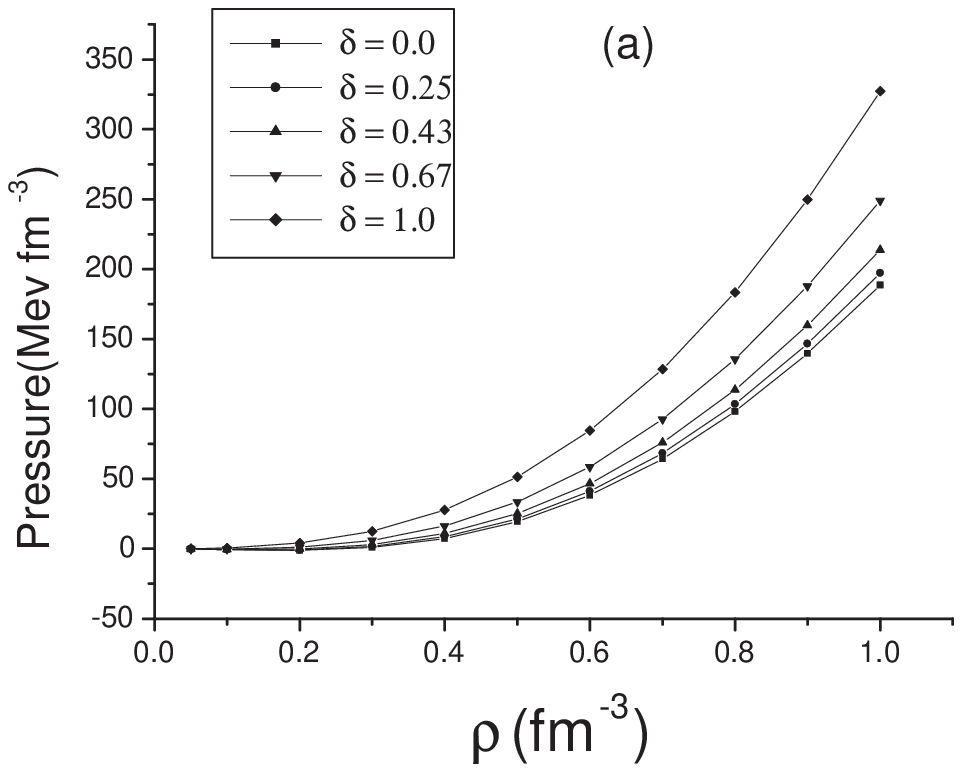}
\includegraphics[height=2.5in]{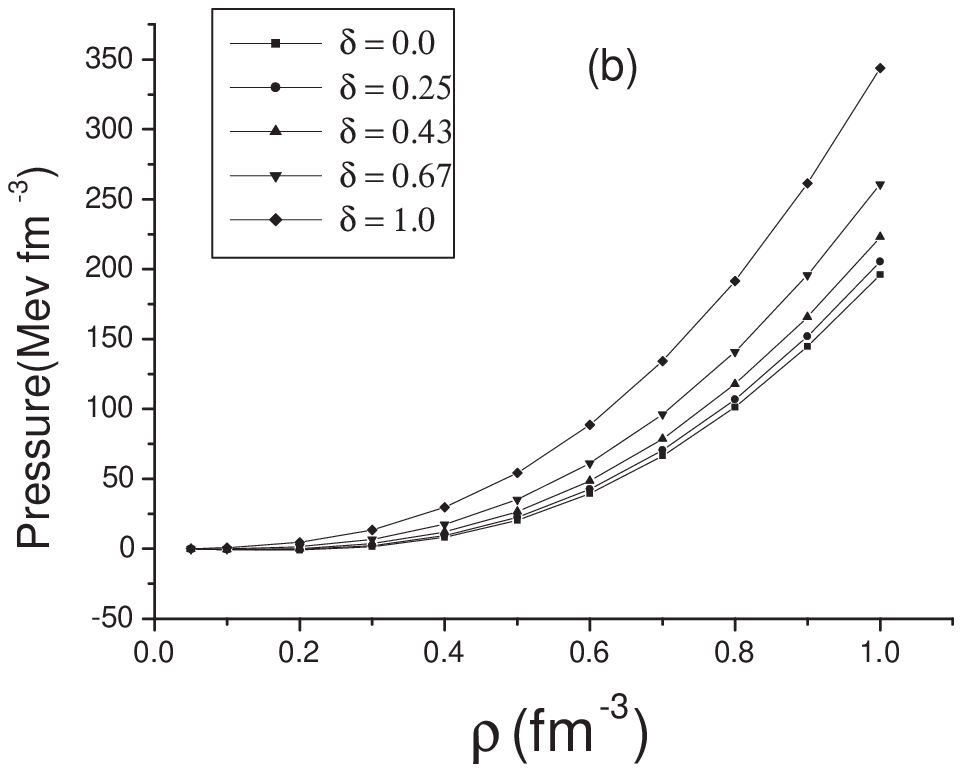}
\includegraphics[height=2.5in]{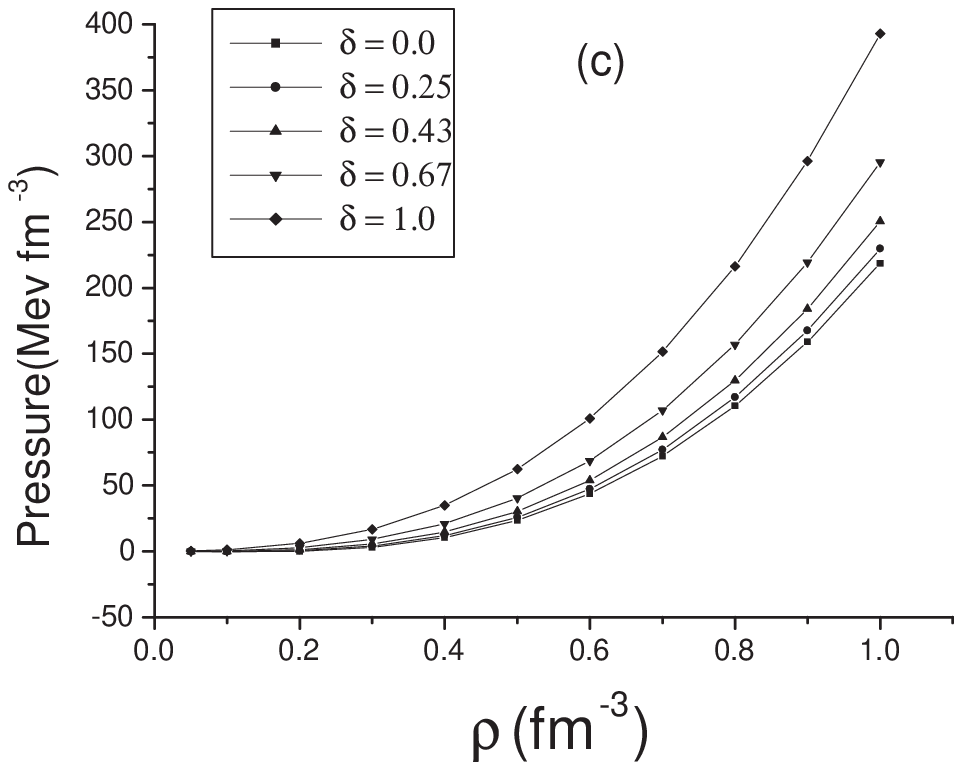}
\includegraphics[height=2.5in]{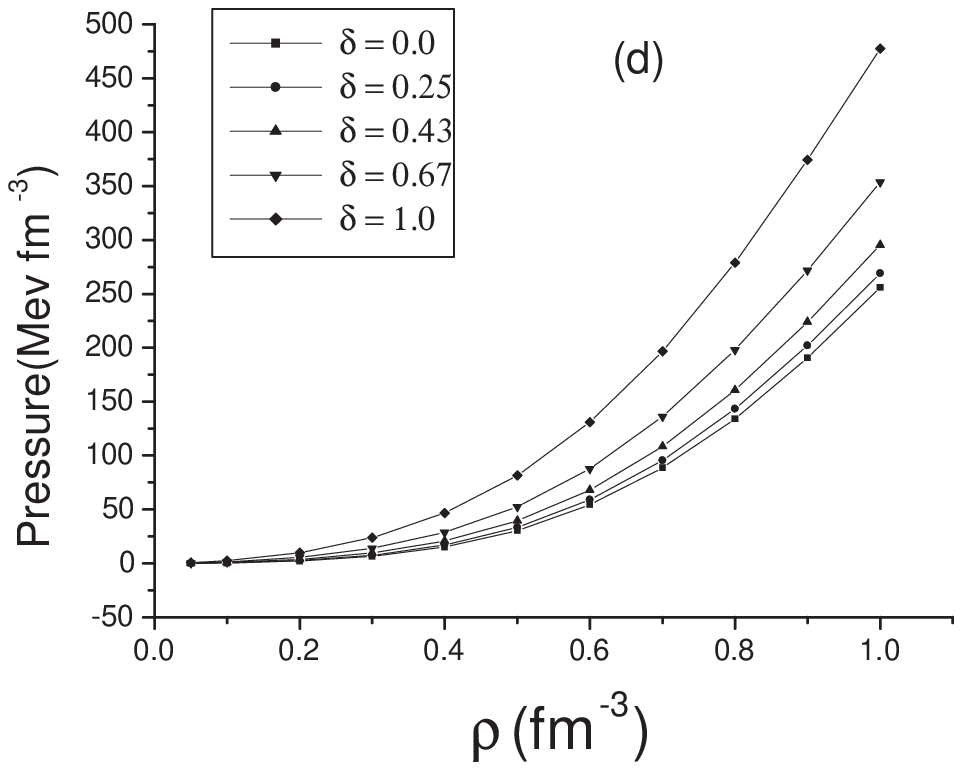}
\caption{The equation of state of polarized asymmetrical nuclear
matter for different values of the spin polarization ($\delta$)
with the $AV_{18}$ potential  at $\beta=0.0$ (a), $\beta=0.3$ (b),
$\beta=0.6$ (c), $\beta=1.0$ (d).} \label{PRE(den)}
\end{figure}

\newpage

\begin{figure}
\includegraphics[height=2.5in]{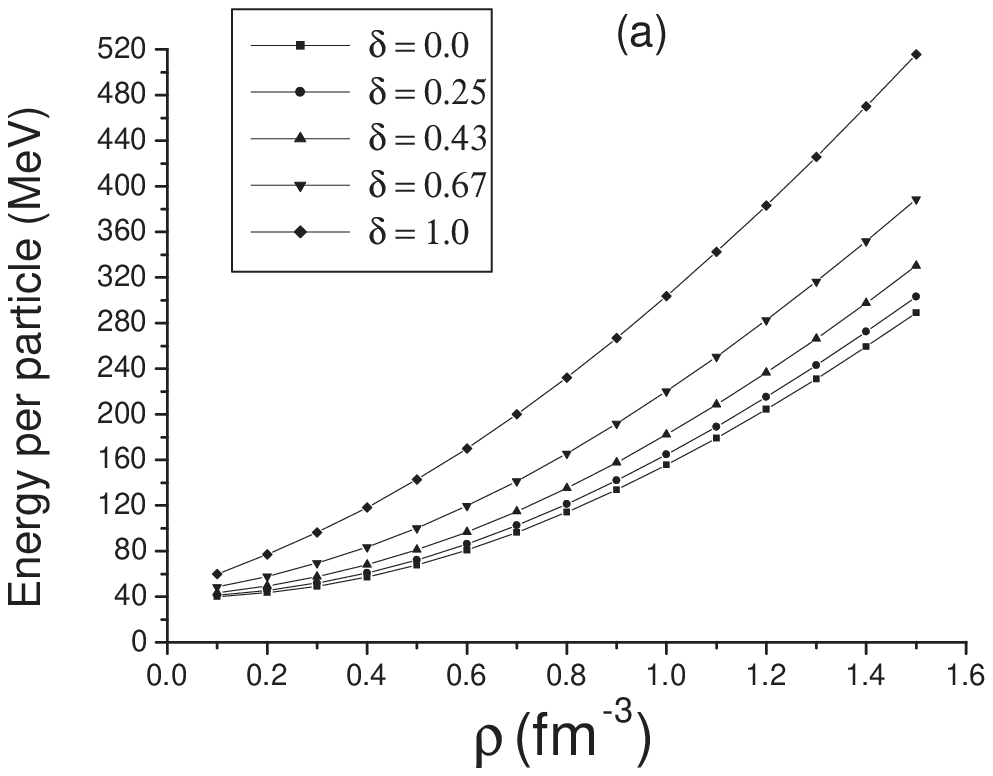}
\includegraphics[height=2.5in]{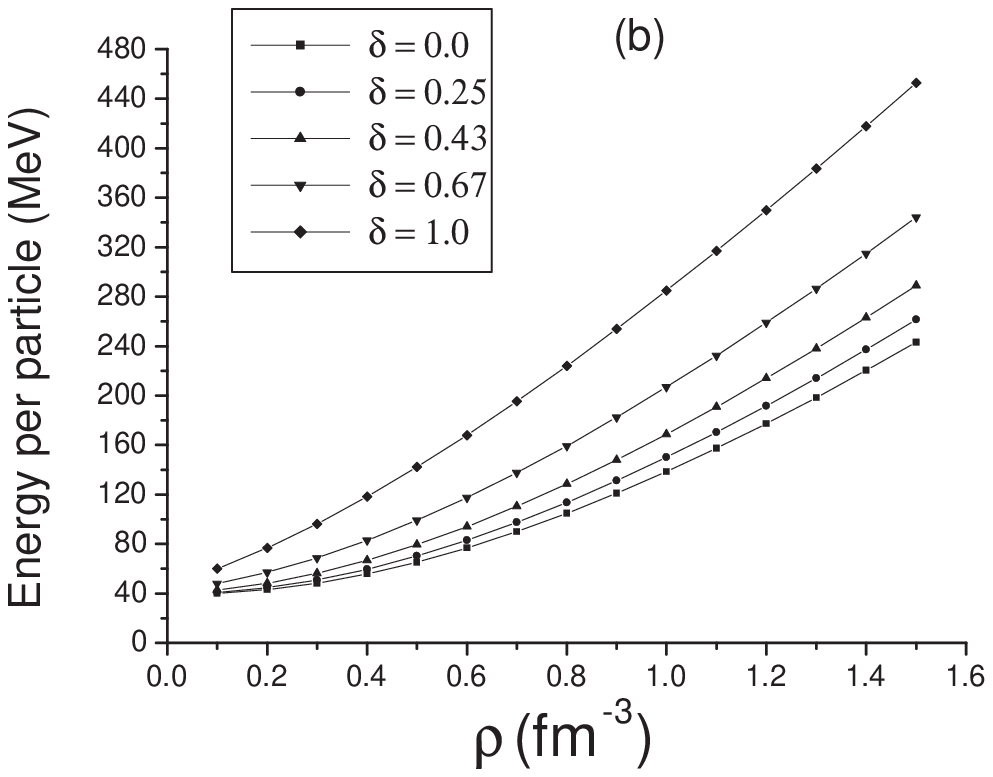}
\includegraphics[height=2.5in]{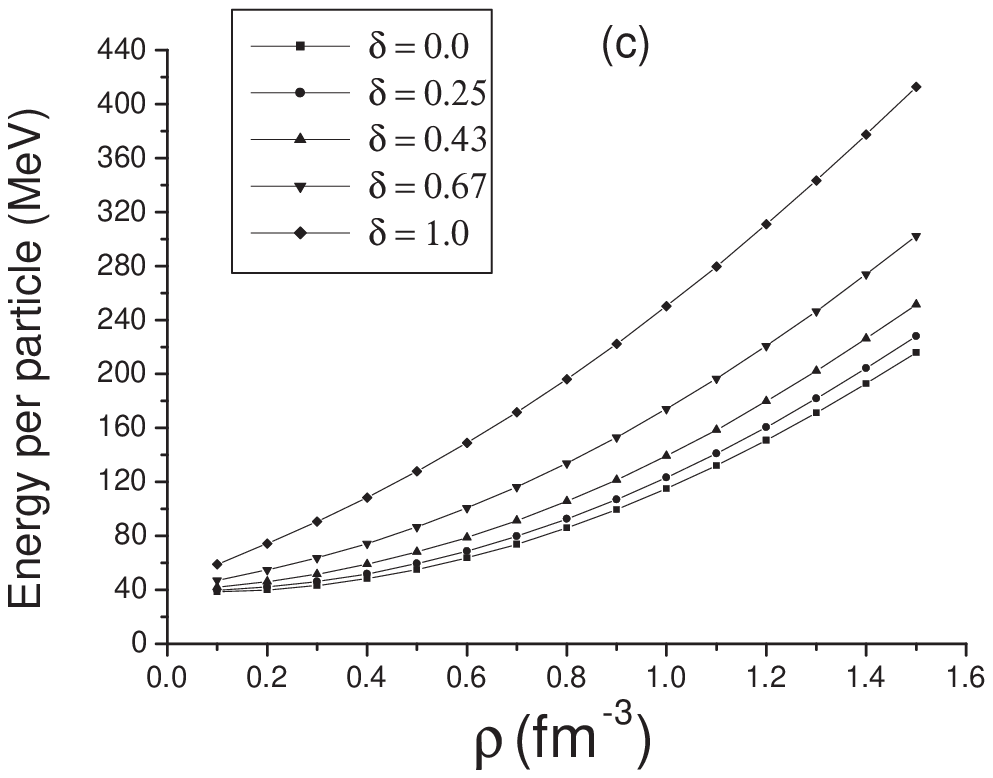}
\includegraphics[height=2.5in]{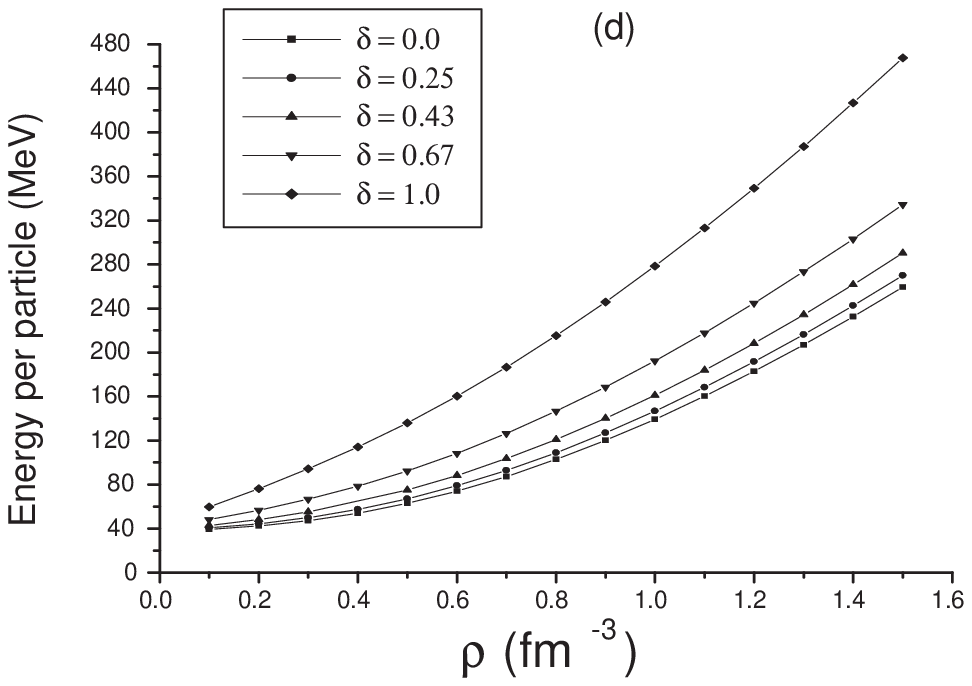}
\caption{The energy per particle of the polarized neutron star
matter versus density ($\rho$) for different values of the spin
polarization ($\delta$) with the $AV_{18}$ (a), Reid93 (b),
$UV_{14}$ (c) and $AV_{14}$ (d) potentials.} \label{ensm)}
\end{figure}

%%%%%%%%%%%%%%%%%%%%%%%%%%%%%%%%%%%%%%%%%%%%%%%%%%%%%%%%%%%%%%%%%%%%%%%%%%%%

\begin{figure} \includegraphics{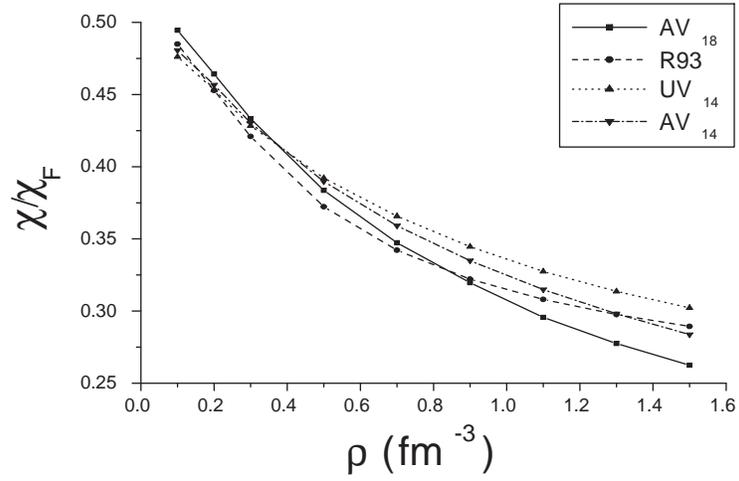}
 \caption{The magnetic susceptibility of the
polarized neutron star matter versus density ($\rho$) with the
$AV_{18}$, Reid93, $UV_{14}$ and $AV_{14}$ potentials. }
\label{sus(den)}
\end{figure}

%%%%%%%%%%%%%%%%%%%%%%%%%%%%%%%%%%%%%%%%%%%%%%%%%%%%%%%%%%%%%%%%%%%%%%%%%%%%%%%%%%%%%
\begin{figure}
\includegraphics{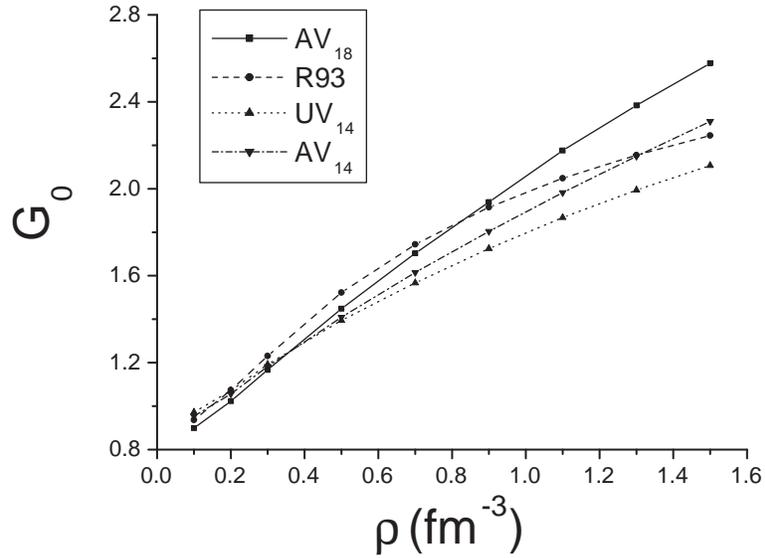} \caption{The
Landau parameter, $G_{0}$, of polarized neutron star matter as
function of density($\rho$) with the $AV_{18}$, Reid93, $UV_{14}$
and $AV_{14}$ potentials.} \label{G(den)}
\end{figure}

%%%%%%%%%%%%%%%%%%%%%%%%%%%%%%%%%%%%%%%%%%%%%%%%%%%%%%%%%%%%%%%%%%%%%%%%%%%%%%%%%%%%
\begin{figure} \includegraphics{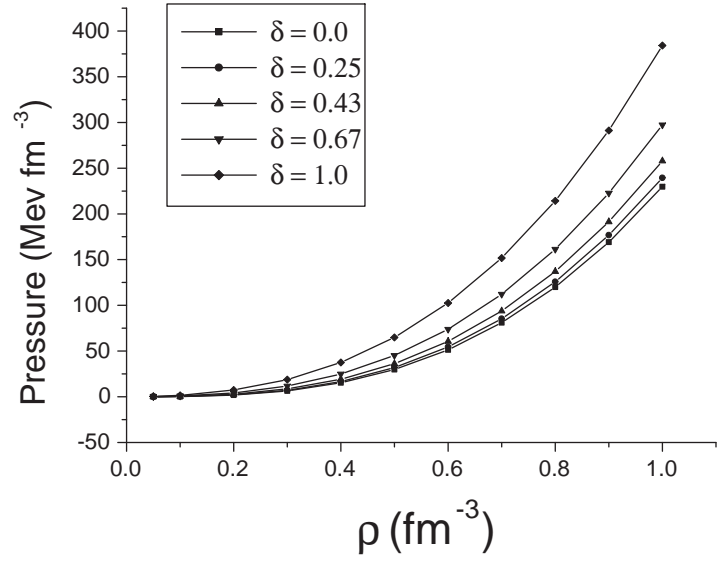}
 \caption{The
equation of state of polarized neutron star matter for different
values of the spin polarization ($\delta$) with the $AV_{18}$
potential .} \label{ensm)}
\end{figure}

%%%%%%%%%%%%%%%%%%%%%%%%%%%%%%%%%%%%%%%%%%%%%%%%%%%%%%%%%%%%%%%%%%%%%%%%%%%%%%%%%%%%%%%%

\end{document}